\documentclass[aps,pre,twocolumn,superscriptaddress,showpacs,preprintnumbers]{revtex4-1}
\usepackage{natbib}
\usepackage[english]{babel}
\usepackage[dvips]{graphics}
\usepackage{graphicx,epsfig}
\usepackage{amsmath}
\usepackage{amssymb}
\usepackage{color}
\usepackage{multirow}
\usepackage[normalem]{ulem}
\usepackage{booktabs}
\usepackage{physics}
\usepackage{dsfont}
\usepackage{float}
\usepackage{amsfonts}
\usepackage{eurosym}
\usepackage{braket}
\usepackage[makeroom]{cancel}
\usepackage[toc,page]{appendix}
\usepackage{amsmath}
\usepackage{amsfonts}

\def\Cbb{\mathbb{C}}

\def\HC{\mathcal{H}}
\def\KC{\mathcal{K}}

\def\kp{\ket{\psi}}
\def\kpt{\ket{\psi(t)}}

\begin{document}
\title{Quantum reservoir complexity by Krylov evolution approach    
}
\author{Laia Domingo}
\email[E--mail address: ]{laia.domingo@icmat.es}
\affiliation{Departamento de Química; Universidad Autónoma de Madrid;
CANTOBLANCO - 28049 Madrid, Spain}
\affiliation{Grupo de Sistemas Complejos; Universidad Politécnica de Madrid; 28035 Madrid, Spain}
\affiliation{Instituto de Ciencias Matemáticas (ICMAT); Campus de Cantoblanco; 
Universidad Aut\'onoma de Madrid;
Nicolás Cabrera, 13-15; 28049 Madrid, Spain}

\author{F. Borondo}
\email[Corresponding author e--mail address: ]{f.borondo@uam.es}
\affiliation{Departamento de Química; Universidad Autónoma de Madrid;
CANTOBLANCO - 28049 Madrid, Spain}

\author{Gast\'on Scialchi}
\affiliation{Departamento de Física “J. J. Giambiagi” and IFIBA, FCEyN, Universidad de Buenos Aires, 1428 Buenos Aires, Argentina}

\author{Augusto J. Roncaglia}
\affiliation{Departamento de Física “J. J. Giambiagi” and IFIBA, FCEyN, Universidad de Buenos Aires, 1428 Buenos Aires, Argentina}

\author{Gabriel G.\ Carlo}
\email[E--mail address: ]{carlo@tandar.cnea.gov.ar}
\affiliation{%
Departamento de F\'sica, Comisi\'on Nacional de Energ\'ia At\'omica, Avenida del Libertador 8250,
(C1429BNP) Buenos Aires, Argentina
}%

\author{Diego A. Wisniacki}
\email[E--mail address: ]{wisniacki@df.uba.ar}
\affiliation{Departamento de Física “J. J. Giambiagi” and IFIBA, FCEyN, Universidad de Buenos Aires, 1428 Buenos Aires, Argentina}
\date{\today}

\begin{abstract}
Quantum reservoir computing algorithms recently emerged as a standout approach in the development 
of successful methods for the NISQ era, because of its superb performance and compatibility with current 
quantum devices.  
By harnessing the properties and dynamics of a quantum system, quantum reservoir computing effectively 
uncovers hidden patterns in data.  
However, the design of the quantum reservoir is crucial to this end, in order to ensure an 
optimal performance of the algorithm. 
In this work, we introduce a precise quantitative method, with strong physical foundations based on 
the Krylov evolution, to assess the wanted good performance in machine learning tasks. 
Our results show that the Krylov approach to complexity strongly correlates with quantum reservoir performance,
making it a powerful tool in the quest for optimally designed quantum reservoirs,
which will pave the road to the implementation of successful quantum machine learning methods.
\end{abstract}

\maketitle

\section{Introduction} \label{sec:intro}

Quantum computing is currently a rapidly evolving field, exerting a substantial impact on various domains, including, cryptography, optimization, and machine learning. 
Despite remarkable recent progress, the development of fault-tolerant quantum computers, 
capable of solving most challenging tasks such as integer factorization \cite{Shor} 
or unstructured search \cite{grover}, remains a long-term goal that necessitates extensive error correction for a significant number of qubits.
As an interesting alternative, noisy intermediate-scale quantum (NISQ) \cite{NISQ} algorithms, 
which leverage the current generation of quantum computers with tens or hundreds of qubits, 
have recently accomplished important milestones \cite{Preskill18}. 

A promising application of NISQ devices is quantum machine learning.
In particular, Quantum Reservoir Computing (QRC) \cite{Fujii2021,QRC2} has emerged as a powerful 
algorithm, demonstrating an excellent performance on a wide range of applications 
\cite{DynamicalIsing, OptQRC, CITA, domingo2023taking}.
Notably, QRC represents a significant leap in the realm of machine learning, built upon the foundations of classical reservoir computing \cite{ESN}. 
While classical reservoir computing relies on the dynamic properties of neural networks for computational tasks, 
QRC exploits the dynamics and properties of characteristic quantum systems as computational resources to perform 
machine learning tasks. 
These dynamics have been recently shown to play a key role in the efficiency of QRC \cite{DynamicalIsing},
and the main goal of this work is to provide an adequate venue for its optimal design, 
that allows the processing of larger input datasets taking advantage of the exponential size of the 
associated Hilbert space. 
Within this framework, QRC has been able to solve classification \cite{imageQRC},
regression \cite{CITA, quantumchemQRC}, and forecasting temporal tasks 
\cite{zambrini_npj, zambrini_com_phys, time_series2, time_series3,time_Series4,time_series5, OptQRC,QRC2}.

In gate-based quantum computing, QRC relies on a random quantum circuit, 
aka the \emph{quantum reservoir}, 
which is applied to an initial quantum state representing the input data. 
The objective is to extract valuable information from it by measuring local operators, 
yielding the relevant features necessary for predicting the output,
which are then fed into a classical machine learning algorithm, typically a linear model. 
It has been shown that, for an effective learning of input-output relationships, 
the quantum reservoir must be a complex enough quantum circuit 
\cite{CITA, domingo2023taking, DynamicalIsing}.
Hence, a careful design of the quantum reservoir is vital to achieve an optimal 
performance of the QRC model. 
%
The majorization principle \cite{majorization_original}, a statistical concept used to assess 
the degree of discordance between probability distributions, 
has emerged as a significant and efficient indicator of complexity for random quantum circuits
\cite{majorization,latorre2002majorization,vallejos2021principle}, and it has proven to be a 
compelling indicator of performance in QRC as well \cite{CITA}. 
However, given the recent advancements in quantum technologies, there is a pressing need 
for more powerful complexity methods to understand the physical properties of 
quantum algorithms that can facilitate the optimal design of large-scale quantum solutions. 

In this paper, we present a significant step forward along this line, by delving into a novel 
complexity measure known as the Krylov complexity and the associated Lanczos sequence
\cite{Parker2019,rabinovici2021operator}, which have been proposed  
for operators \cite{PhysRevX.9.041017} 
and states \cite{balasubramanian2022quantum}.
These measures are built on robust physical principles, and the Krylov complexity's 
unique growth behavior provides a precise definition of complexity over large time scales.
Moreover, they offer applicability to a broad variety of problems, 
having proven suitable for analyzing intricate systems in condensed matter physics
\cite{Ballar_Trigueros2022}, quantum field theory \cite{KryQFT}, and quantum information \cite{KryQIT}.
In this sense, our analysis of the Krylov complexity paves the road to unlock deeper insights 
into quantum systems, facilitating the development of superior quantum algorithms for different 
real-world challenges.

Originally developed to efficiently calculate the exponential of a matrix 
\cite{Hochbruck1997, Parlett1998}, Krylov methods have found widespread use in quantum evolution 
of states and operators within systems with large Hilbert spaces \cite{Ruffinelli2022}.
These methods involve mapping the system's evolution to a non-interacting tight-binding model 
within the so-called Krylov space. 
Such mapping leads to a measure of complexity that has garnered significant attention 
in recent research
\cite{Dymarsky2020,Ballar_Trigueros2022,Cao2021,Bhattacharjee2022_2,Bhattacharya2022,
Bhattacharjee2022,PhysRevE.107.024217,PhysRevX.9.041017,barbon2019evolution,rabinovici2021operator,
Rabinovici_2022_localization,Rabinovici2021,Rabinovici_sup2022}. 
The key aspect of this measurement, called K-complexity, lies in the concept of spreading of the 
one-particle wavefunction over Krylov basis. 
In fact, it has been demonstrated that, by using this basis, the dispersion of a wave packet is minimized, thereby avoiding ambiguities when defining the complexity associated with quantum evolution.
Several authors have extensively considered this measure to study the transition from integrability 
to chaos in a variety of systems \cite{Rabinovici_2022_integrability,camargo2023spectral,hashimoto2023krylov}.
It is important to note that when applied to operators, the effectiveness of this complexity measure 
appears to be contingent on the specific operator under consideration \cite{PhysRevE.107.024217}. 
Similarly, in the case of states, it may not yield
satisfactory results depending on the initial 
conditions \cite{Diego}. 
However, in the case of state evolution, and within the effective tight-binding model,
both the Krylov complexity and the statistics of the onsite potentials have shown to be promising 
robust measures of quantum complexity.

In this paper, we focus on the investigation of the reservoir complexity using Krylov methods, 
to advance in the optimal design for QRC. 
However, an important challenge arises in the case of reservoirs, as we lack a Hamiltonian description, 
having instead an evolution operator $U(t)$.
To overcome this hurdle, we propose to use an effective Hamiltonian for these reservoir systems, 
which involves taking the logarithm of the operator $U(t)$. 
To validate this approach, we first delve into analyzing spin chains, as a starting point of our research. 
By careful examination of the system behavior at different time instances, 
the Krylov approach is found to effectively quantify the system complexity only when the evolution 
operator is used for times smaller than the \emph{scrambling time} \cite{García-Mata:2023},
defined as the point where the K-complexity of the system reaches its asymptotic plateau. 
Beyond this time, the system exhibits a level of chaoticity that makes it challenging to describe 
it using conventional methods.
We then further proceed with our validation by analyzing the complexity of the standard map, 
the paradigmatic system extensively studied for both classical and quantum chaos, where,
as stated before, the quantum dynamics is governed by an evolution operator,
without a Hamiltonian description. 
Our results demonstrate that the Krylov complexity aligns perfectly with other conventional 
measures of quantum chaos.
Most relevant, by employing the Krylov-based complexity measure, 
a valuable insight into the intricate dynamics of quantum reservoirs is gained, 
especially for times before the onset of significant scrambling effects. 
Additionally, a pivotal objective of our research is to shed light on the relationship 
between Krylov  methods and other well-established measures of complexity, providing
a comprehensive understanding of these intriguing quantum systems. 


\section{Results} \label{sec:results}

We next analyze the performance of quantum reservoirs 
using the Krylov approach to complexity. 

In the realm of QRC, random quantum circuits are represented by a unitary evolution operator $U$, 
which creates a quantum-enhanced representation of the input data. 
A problem arises here;
although the Krylov method is typically well-defined for studying quantum systems when described 
by a Hamiltonian operator, there is no satisfactory adaptation 
for the case of unitary operators. 
This severe limitation needs to be addressed beforehand,
to make it possible to study the Krylov complexity for quantum reservoirs.
To this end, we introduce in Sec.~\ref{sec:solution} a suitable method to analyze the Krylov 
complexity with quantum unitaries, 
which is then validated by showing that it can accurately reproduce the properties 
of two benchmark quantum systems with chaotic dynamics: the longitudinal-transverse field Ising model, 
and the standard map. 

\subsection{Krylov complexity for evolution operators} \label{sec:solution}
The method that we propose to use is simple.
Given an evolution operator $U$, a corresponding effective Hamiltonian $H_\text{eff}$ is calculated as
\begin{equation}
    H_\text{eff} = -i \log U.
    \label{eq:H_eff}
\end{equation}
We will show that this choice allows us to use the above method to characterize the $K$-complexity of $U$.
The operator $H_\text{eff}$ is then used to calculate the Lanczos coefficients and Krylov complexity 
for an initial quantum state $\ket{\psi}$. 



To validate this ansatz, we first apply the proposed method to the longitudinal-transverse field Ising model, 
with nearest neighbor interactions in the $z$ direction, and a transverse magnetic field in the $(x, z)$ plane.
The corresponding Hamiltonian is then given by
\begin{equation}
    H = \sum_{k=1}^n (h_x \sigma_k^x + h_z \sigma_k^z) - J \sum_{k=1}^{n-1} \sigma_k^z\sigma_{k+1}^z,
    \label{eq:Ising}
\end{equation}
where $n$ is the number of spin $-1/2$ sites in the chain, $\sigma_k^j$ is the Pauli operator at site 
$k = 1, 2, \cdots , n$ in the $x, y$ and $z$ directions, $h_x$ and $h_z$ are the components of the 
magnetic field, and $J$ is the nearest-neighbor coupling. 
To expand an operator in the Krylov space, it is necessary to work in a symmetry subspace. 
Our system is invariant under reflection with respect to the centre of the chain. 
In this work, we always operate in the positive parity subspace and fix the coefficients $h_x=J=1$. 
Hamiltonian (\ref{eq:Ising}) is integrable for $h_z=0$ and for $h_z \gtrsim 3$, while it exhibits a 
quantum chaotic behavior for intermediate values of $h_z$. 

\begin{figure}
    \centering
    \includegraphics[width=0.85\columnwidth]{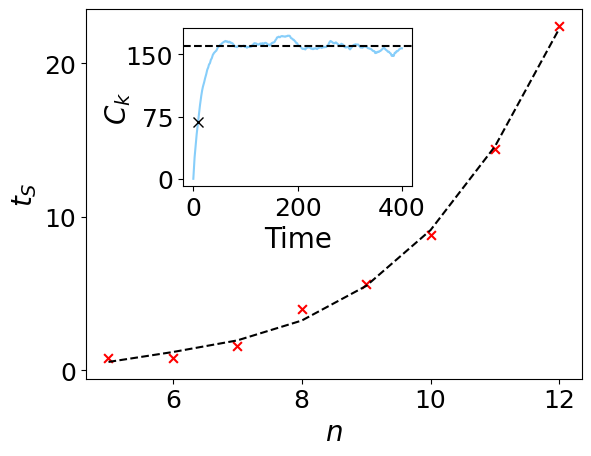}
    \caption{Scaling of the scrambling time ($t_S$) with the number of quantum spins ($n$) in the 
    Ising model (\ref{eq:Ising}). 
    The inset illustrates the calculation of the scrambling time, defined as the duration required 
    for the $K$-complexity $C_k$, as defined in Eq.~(\ref{eq:K_complexity}) to reach half of the 
    average value at the plateau for a system of size $n=10$.}
    \label{fig:scrambling-time}
\end{figure}

We next construct the evolution operator $U = e^{i H T/\hbar}$ for our Hamiltonian. 
In this case, the unitary is also normalized to the chosen time $T$, to recover the scale of the original Hamiltonian. 
Finally, the chaotic properties of $H_\text{eff}$ associated with the Ising model are computed for three different values of the characteristic time. 
The longest one is taken as the Heisenberg time, 
\begin{equation}
    t_H = \frac{2 \pi \hbar}{\rho_E},
    \label{eq:Heisenberg}
\end{equation}
where $\rho_E$ represents the mean density of states around energy $E$. 
The second time is the scrambling time $t_S$, defined as the time it takes to reach half the value 
of the saturation of the $K-$complexity calculated using $H$ of Eq.~(\ref{eq:Ising}). 
Figure~\ref{fig:scrambling-time} shows the scaling of the scrambling time with the number of spins of the quantum system, 
which increases cubically with $n$. This power law behavior is characteristic of integrable or mixed systems than chaotic \cite{PhysRevE.74.056208}, which requires a more detailed study that we will do in the future.
A representation of how the scrambling time is defined is shown in the inset. 
Finally, a third time is chosen, corresponding to a short period of time $t_S/25$. 

\begin{figure}
    \centering
    \includegraphics[width = 0.85\columnwidth]{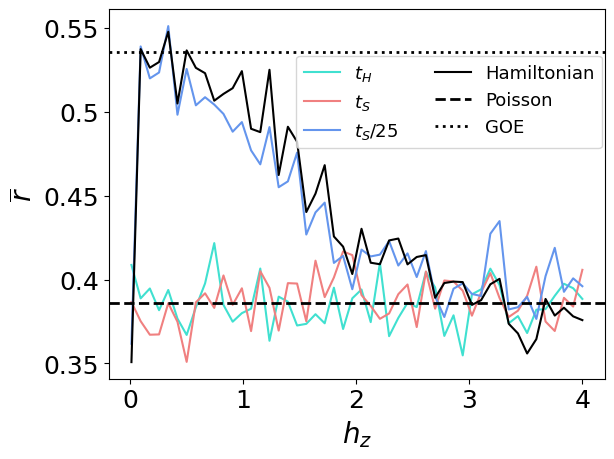}
    \caption{Mean value of the distribution of the ratio of consecutive level spacings $\overline{r}$ for the Ising model  
    of Eq.~(\ref{eq:Ising}) as a function of the coupling constant $h_z$. 
    The values $\overline{r} \approx 0.386$  (horizontal dashed line) and  $\overline{r} \approx 0.536$ (horizontal dotted line)
    corresponds to a regular and a chaotic system, respectively.
    The Hamiltonian has been constructed from the unitary evolution operator for three different times: 
    the Heisenberg time ($t_H$), the scrambling time ($t_S$), and a shorter period of time ($t_S/25$). 
    The black curves represent the statistics obtained with the Ising model Hamiltonian.}
    \label{fig:chaos-H}
\end{figure}

With these three time scales, we study the chaotic dynamics of the Ising Hamiltonian for different 
values of $h_z$ for a system with $n=10$ spins. 
The results are displayed in Fig.~\ref{fig:chaos-H}. 
There, the black curve shows the 
mean value of the distribution of the ratio of consecutive level spacings
$\overline{r}$ calculated 
with the actual Hamiltonian in Eq.~(\ref{eq:Ising}) \cite{oganesyan2007localization,kudo2018finite,atas2013distribution}, which indicates the regular or chaotic 
character of the system according to criteria of random matrix theory \cite{Bohigas1984}. 
This curve clearly shows a very sudden phase transition from an integrable system for $h_z\approx 0$ 
to a chaotic system for $h_z \approx 1$, and gradually back to an integrable system for $h_z \gtrsim 3$. 
On the other hand, the color curves show the values of $\overline{r}$ calculated with the effective 
Hamiltonian $H_\text{eff}$ for the three characteristic times using Eq.~(\ref{eq:H_eff}). 
As can be seen, when the time scale is short, i.e.,~$t_S/25$, $H_\text{eff}$ correctly reproduces 
the behavior of $\overline{r}$. 
However, when the characteristic time increases, the resulting effective Hamiltonian loses the 
phase transition, becoming integrable for all values of $h_z$. 

\begin{figure}
    \centering
    \includegraphics[width = 0.85\columnwidth]{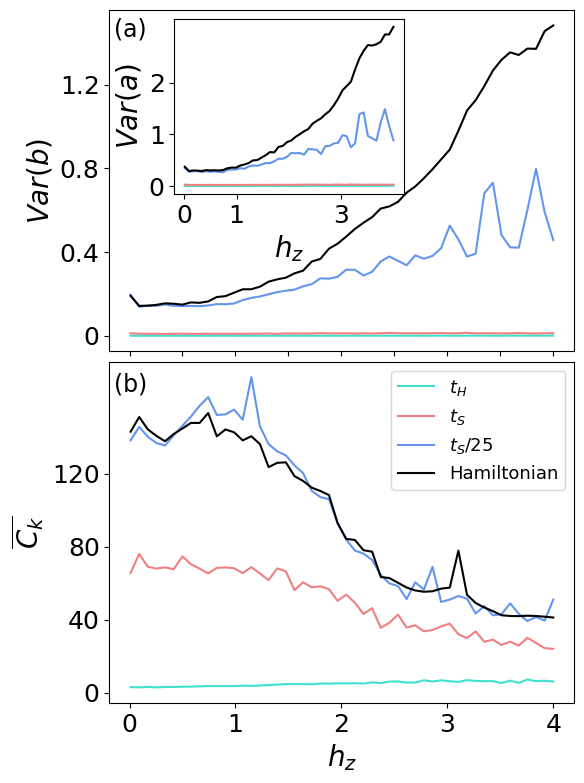}
    \caption{(a) Variance of the Lanczos coefficients $a$ and $b$ for the Ising model with $n=10$ spins, 
    as a function of $h_z$. 
    (b) Mean $K$-complexity ($\overline{C_k}$) as a function of $h_z$. 
    The three times used are the same as in Fig.~\ref{fig:chaos-H}.}
    \label{fig:complexity-H}
\end{figure}

To complete our study, 
we show in Fig.~\ref{fig:complexity-H} the mean $K-$complexity, $\overline{C_k}$, and the variance 
of the Lanczos coefficients $a$ and $b$ for the same three time scales considered before. 
%
The Krylov complexity has been computed using 
several initial states $\ket \psi$ [see Eq.~(\ref{eq:kry_subs})], that are eigenstates of $H$ in the integrable regime ($h_z>5$).
Let us remark that the choice of the initial states is vital to accurately capture the evolution 
of the $K$-complexity with $h_z$ \cite{Diego}.
The results in Fig.~\ref{fig:complexity-H} (a) shows that the variance of $a$ and $b$ increases 
with $h_z$ for the true Hamiltonian (black curve). 
However, as the time period becomes longer, the variance of such coefficients increases 
in a slower fashion (colored curves). 
In fact, in the limit $t = t_H$ the variance of the coefficients is basically constant. 
Complementarily,
Fig.~\ref{fig:complexity-H} (b) shows $\overline{C_k}$ at the saturation time for the different time scales,
from which it can be concluded that,
while for $t=t_S/25$ the dynamics of $\overline{C_k}$ is correctly reproduced, 
the pattern is completely lost at $t=t_H$, and  for $t = t_S$ it exhibits an intermediate dynamics. 

The previous results show that as long as the time of the evolution operator is small enough, 
the effective Hamiltonian, introduced by us in Eq.~(\ref{eq:H_eff}), is able correctly reproduce
the chaotic dynamics of the system ($\overline{r}$), as well as the $K-$complexity ($\overline{C_k}$). 
Notably, the computational time and resources required to apply this method to large-scale 
systems can be substantially reduced by considering \emph{only} the first few 
coefficients of the Lanczos sequence.
As it is shown in the Supplemental Material~\ref{sect:reduced_Lanczos}, 
the coefficients still allow for an accurate reproduction of the Krylov statistics and complexity.

\begin{figure}
    \centering
    \includegraphics[width=.85\columnwidth]{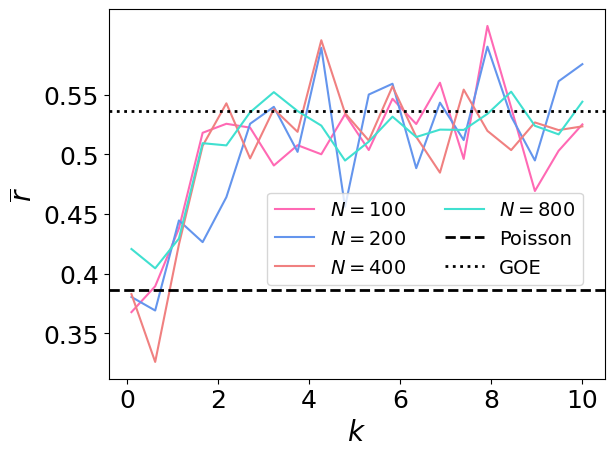}
    \caption{Variation of $\overline{r}$ as a function of the chaos parameter $k$ for the standard map.}
    \label{fig:chaos-U}
\end{figure}


To further confirm our previous conclusions, we consider a second benchmark system,
with different mathematical characteristics.
This system is the (quantum) standard map, a quantum mechanical version of the classical standard map, 
which describes the dynamics of a periodically kicked rotor system on a torus. 
The associated torus structure gives rise to periodicity in both position and momentum. 
When quantized, this periodicity leads to a discrete Hilbert space with a dimension of $N$, 
and Planck constant denoted as $h=1/(2\pi N)$.

Mathematically, the standard map is defined by a unitary operator $U$, such that
\begin{equation}
    \ket{\psi} \longrightarrow U \ket{\psi} = e^{ -i \; \frac{{\hat p}^2}{2\hbar}} \;
         e^{ -i \; \frac{k}{\hbar} \cos(2 \pi \hat x) } \ket{\psi},
\end{equation}
where $\hat x$ is the position operator, $\hat p$ is the momentum operator, and $k$ is the chaos parameter. 
For small values of $k$, the level spacing statistics ($\overline{r}$) is described by the Poisson law, 
and by the Wigner-Dyson law of the random matrix theory for large values of $k$. 
This behavior is depicted in Fig.~\ref{fig:chaos-U}, which shows the variation of $\overline{r}$ 
with the parameter $k$ for different sizes of the Hilbert space. 
As can be seen, the evolution of $\overline{r}$ is qualitatively equivalent for all different system 
sizes considered.

\begin{figure}[t]
    \centering
    \includegraphics[width=0.85\columnwidth]{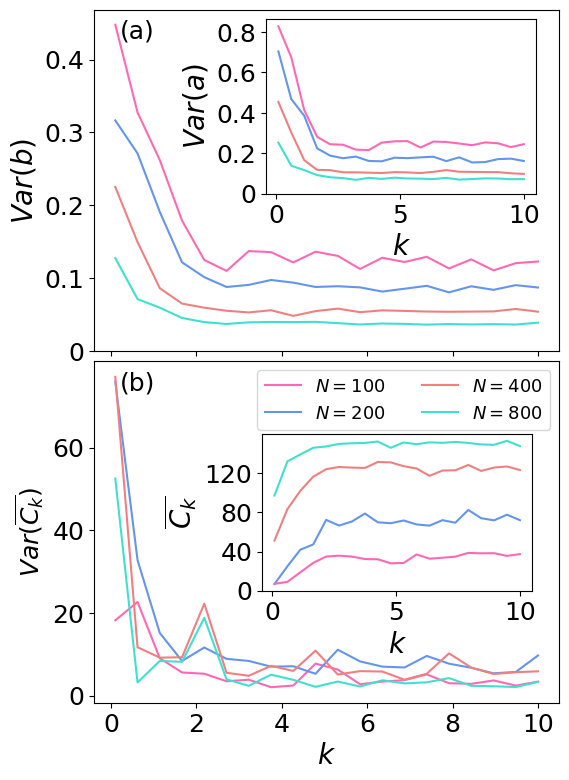}
    \caption{Same as Fig.~\ref{fig:complexity-H} for the standard map with Hilbert space sizes 
    $N=100, 200, 400$ and $800$.}
    \label{fig:complexity-U}
\end{figure}

Similarly to what we did before with the Ising model, for which the system Hamiltonian was known, we have obtained the mean complexity $\overline{C_k}$ and variance of the Lanczos 
coefficients with the effective Hamiltonian $H_\text{eff}$ in Eq.~(\ref{eq:H_eff}). 
The results, displayed in Fig.~\ref{fig:complexity-U}, show that also in this case the Krylov statistics 
correctly reproduce the quantum chaotic behavior of the system. 
In particular, Fig.~\ref{fig:complexity-U} (a) shows that the variance of the Lanczos coefficients 
decreases as $k$ increases, in accordance with the behavior of the mean level spacing  $\overline{r}$. 
Moreover, Fig.~\ref{fig:complexity-U} (b) shows that the mean and variance of the $K-$complexity 
also increase (decrease) with $k$, result which also reproduces the integrable-chaotic transition 
taking place in the system with $k$. 
Here, variances are calculated over the 100 simulations varying the initial state $\ket{\psi}$ of the system (see Eq. \ref{eq:kry_subs}), which are all chosen to be eigenstates of the standard map in the integrable region with $k=0.01$.

In summary, we can state that our results reflect that the Krylov dynamics 
obtained with the effective Hamiltonian correctly captures the complexity and chaotic properties 
of a quantum map. 
This confirms the fact that our construction of $H_\text{eff}$ is a valid method to compute 
the $K-$complexity of a unitary operator, which will be next used to calculate the Krylov statistics 
for the title case of multiple quantum reservoirs. 

\subsection{Krylov complexity for quantum reservoirs}

\begin{figure}
    \centering
    \includegraphics[width=0.85\columnwidth]{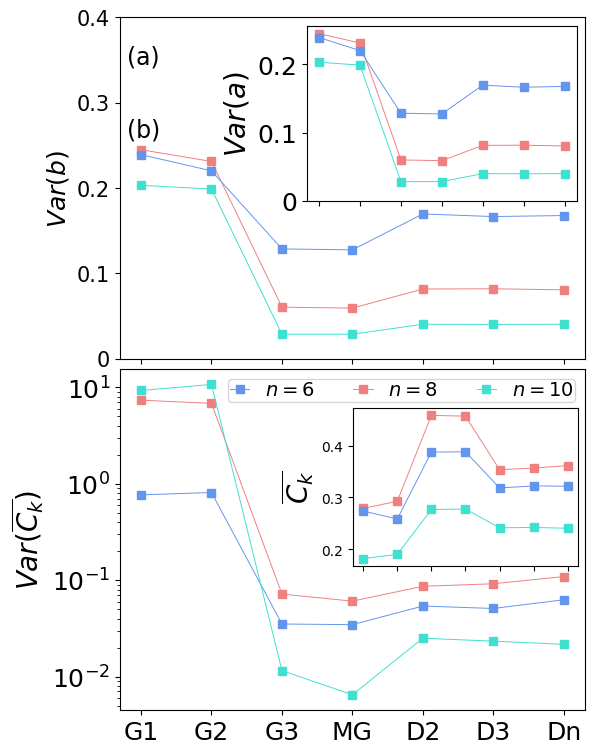}
    \caption{Same as Fig. \ref{fig:complexity-H} for the seven families of quantum reservoirs for 
    $n=10,8$ and $6$.
$\overline{C_k}$ and $Var(\overline{C_k})$ have been normalized with the dimension of the Hilbert space for better visualization.}
    \label{fig:complexity-QR}
\end{figure}

Once our method, described in Sect.~\ref{sec:solution}, has been validated, we proceed to discuss the 
results for the chaotic and Krylov statistics of seven families of random quantum circuits, 
which often serve as quantum reservoirs, and compare them with their performance in QRC 
(see Sect.~\ref{sec:MLtask}). 

We begin by analyzing the statistics of the Lanczos coefficients, together with the mean and variance 
of the $K-$complexity for the seven families of quantum reservoirs. 
The corresponding results are presented in Fig.~\ref{fig:complexity-QR}. 
In particular, Fig.~\ref{fig:complexity-QR} (a) depicts the variance of the Lanczos coefficients 
$a$ and $b$, showing that the families achieving better performance (lower MSE) are also those 
with smaller variance in $a$ and $b$. 
The variances here have been calculated over 100 simulations performed for each family. 
The initial state $\ket{\psi}$ is here chosen randomly from the computational basis.
Moreover, Fig.~\ref{fig:complexity-QR} (b) also indicates the existence of a significant correlation 
in the mean and variance of the $K-$complexity, with the better-performing reservoirs having lower
Var($\overline{C_k}$) and higher $\overline{C_k}$. 
Therefore, these results clearly show that the Krylov statistics are an excellent choice for designing optimal quantum reservoirs, that present high performance for quantum machine learning tasks.

One final point is worth discussing here.
In Refs.~\cite{CITA, domingo2023taking},  
we studied the distribution of the unitary operators generated by the seven 
families of quantum reservoirs in the Pauli space, which is the space of quantum operators. 
Using a simple toy model of two qubits, we demonstrated 
that the less complex families of reservoirs, G1 and G2, only span a small subspace of the operator space. 
On the other hand, the more complex families, such as G3 and MG, uniformly sample the entire operator space,
resulting in improved performance in QRC.
Now, the Krylov approach to complexity further clarifies and allows a better understanding of this result.
Indeed, the $K-$complexity measures the average dimension of the Krylov subspace required 
to represent the evolution of an initial state over time, described by the unitary operator 
sampled from any of the quantum reservoir families. 
Therefore, only the families able to generate the whole set of quantum operators will be able 
to span larger Krylov subspaces, then giving rise to higher values of $\overline{C_k}$. 
On the contrary, families that only produce non-universal, limited sets of quantum operators 
will result in low-dimensional Krylov spaces, leading to lower complexity values.

\begin{figure}
    \centering
    \includegraphics[width=0.85\columnwidth]{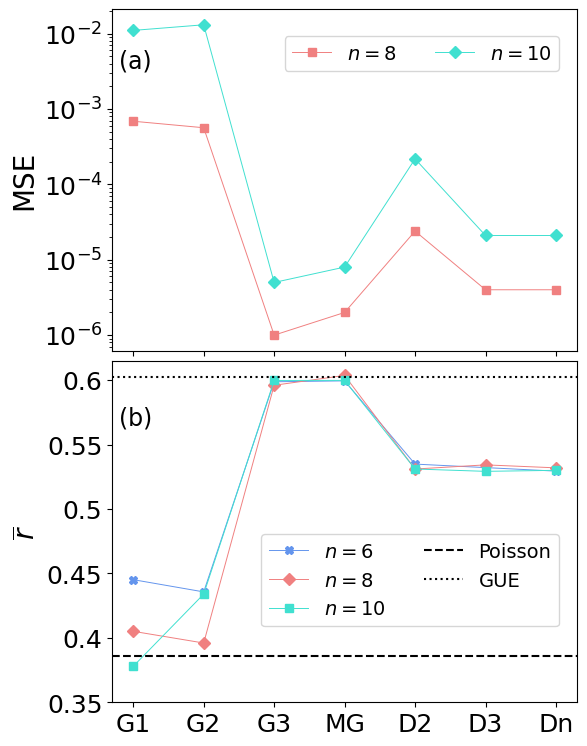}
    \caption{(a) Mean Squared Error (MSE) obtained with the quantum reservoir computing algorithm 
    for the seven families of quantum reservoirs and the two molecules: LiH ($n=8$) and H$_2$O ($n=10$). (b) Variation of $\overline{r}$ for the different 
    families of quantum reservoirs with sizes $n=6,8$ and $10$.
    }
    \label{fig:mse-chaos}
\end{figure}

This is clearly seen in the results of Fig.~\ref{fig:mse-chaos}, which compares the mean squared error 
(MSE) of the predictions in the machine learning tasks [panel (a)] reported in Ref.~\cite{CITA}, 
with the mean energy level spacing, $\overline{r}$, for the seven families 
of quantum reservoirs [panel (b)] obtained here. 
The results show that there is a strong correlation between the quantum chaotic behavior of the families 
of quantum circuits and the performance in QRC. 
Actually, the families with higher values $\overline{r}$, such as the G3 and MG, 
are also the ones providing better performance as quantum reservoirs. 
On the contrary, G1 and G2 are the families with worse performance, which also present lower 
values of $\overline{r}$. 
Finally, the diagonal circuits $D_2$, $D_3$ and $D_N$ provide intermediate results both in terms 
of MSE and  $\overline{r}$.

\section{Methods}\label{sec:Method}

\subsection{Quantum reservoir computing}

The idea of QRC lies in using a Hilbert space as a  high-dimensional embedding of the input data. 
In this way, the extracted features enhanced by quantum operations are used to feed a classical 
machine learning model, which predicts the desired target.  

Consider the dataset $\{(x_i,y_i)\}$, where $x_i$ are the input samples and $y_i$ the target outputs. 
The goal of the machine learning algorithm is to predict the desired output $y_i$ given an input data sample $x_i$. 
In this work, the input data $x_i$ represents the electronic ground state of a molecule, 
and the output $y_i$ is the first excited energy of such molecule (see Sect. \ref{sec:MLtask}). 
The data samples are encoded as an $n$-qubit quantum state $\ket{x_i}$. 
Then, a \emph{random} unitary transformation $U$, implemented by a quantum circuit, 
is applied to extract features from the input data, resulting in the quantum state $U \ket{x_i}$. 
The operator $U$ is sampled from a carefully selected family of operators, such that $U$ creates 
enough entanglement to generate useful transformations of the input data while being experimentally feasible.
For this reason, the design of the reservoir $U$ is crucial for the optimal performance of the algorithm. 
In this work, the QRs are designed as random quantum circuits whose gates are chosen from a family 
containing a finite set of gates. That is, given a family of quantum gates, a quantum circuit is generated by uniformly sampling gates from that family. The depth of the circuit is fixed to $\text{depth} = 40$, and 100 simulations are run for each of the families. We have repeated the experiments with circuits with different number of qubits, $n=6,8$ and $10$. The $K-$complexity and the statistics of the Lanczos coefficients of the resulting unitary operators $U$ will be compared with the performance of the quantum reservoirs in the machine learning task presented in Sect. \ref{sec:MLtask} for two molecules: LiH, requiring $n=8$ qubits, and H$_2$O, requiring $n=10$ qubits.

 After applying $U$ to the initial quantum state, the expected value of single-qubit observables is measured.
 These observables are Pauli operators $\{X_0, Z_0, \cdots X_j, Z_j, \cdots , X_N, Z_N\}$, 
 where $X_j, Z_j$ represent the Pauli operators $X,Z$ applied to qubit $j$. 
 Notice that, in general, a $n$-qubit unitary $U$ transforms a simple observable $Z$ (or $X$) 
 into a linear combination of Pauli operators
\begin{equation}
    UZU^\dag = \sum_k \alpha_k P_k,
\end{equation}
where $\{P_k\}$ are tensor products of local Pauli operators. 
Therefore, measuring single Pauli operators of a state which has received a unitary transformation 
could produce complex nonlinear outputs, which could be represented as a linear combination of 
exponentially many nonlinear functions \cite{QRC}. 

Finally, the extracted features $\hat{x_i}= (\expval{X_1}, \expval{Z_1}, \cdots,\expval{X_N}, \expval{Z_N})$, 
are fed to a classical ML algorithm, usually a linear model. 
Even though more complex models can be used, the quantum reservoir should be able to extract valuable 
features so that a simple machine learning model can predict the target $y$. 
A linear model with regularization, called ridge regression, is enough to learn the output. 
Ridge regression minimizes the following expression
\begin{equation}
   \text{MSE}_r(\hat{y}, y) = \frac{1}{t} \sum_{t=1}^T \left[(y(t) - W \hat{x}(t)\right]^2 
       + \gamma \; ||W||^2 ,
 \label{eq:ridge}
\end{equation}
where $\gamma$ is the regularization parameter and $W$ are the linear coefficients and $||\cdot ||^2$ is the $L^2$ norm. 
This loss function prevents the algorithm from learning large coefficients $W$, 
which usually leads to unstable training and poor generalization capacity. 
$\gamma$ is a hyperparameter which needs to be tuned depending on the problem at hand. 
When $\gamma$ is too large, the model will learn very small values of $W$, 
which leads to predicting constant values. 
On the other hand, if $\gamma$ is too small the chances of overfitting increase.

\subsubsection{Families of quantum reservoirs}
\label{sec:familiesQR}

The design of the quantum reservoir is crucial for the performance of the algorithm. 
For this reason, seven \textit{families} of quantum circuits are studied. For a given family, the quantum circuit is built by adding a fixed number of random quantum gates from such family. For each family, 100 simulations are carried out. 

The seven families of quantum reservoirs can be ordered in terms of complexity according to the majorization principle \cite{majorization}, which in turn corresponds to different performances in QRC \cite{CITA}. 
In this study, we will demonstrate that the distinction in complexity can also be quantified using 
the Krylov evolution. 
As a result, Krylov measures offer excellent means to identify optimal families of random quantum circuits 
for machine learning tasks. 
This approach provides a more precise and quantitative method for comprehending the variations between various quantum reservoir designs, enabling the formulation of practical guidelines for efficient quantum algorithm implementations. 
Additionally, it bridges the gap between quantum machine learning designs and quantum complexity theory, opening up new possibilities for advancements in quantum machine learning research.

The seven families considered are the following. 
The first 3 circuits are constructed from a few generators: 
G1 = \{CNOT, H, X\}, G2 = \{CNOT, H, S\}, and G3=\{CNOT, H, T\}, 
where CNOT is the controlled-NOT gate, 
H stands for Hadamard, 
and S and T are $\pi/4$ and $\pi/8$ phase gates, respectively. 
The circuits constructed from G2 generate the Clifford group \cite{G2}, 
and G1 generate a subgroup of Clifford \cite{G1}. 
Therefore, both G1 and G2 are non-universal and classically simulatable. 
On the other hand, G3 is universal and thus approximates the full unitary group $U(N)$ 
to arbitrary precision. 
The fourth family is composed of Matchgates (MG), which are two-qubit gates formed 
from 2 one-qubit gates, A and B, with the same determinant. Matchgates circuits are also universal (except when acting on nearest neighbor lines only) \cite{matchgates1, matchgates2}. 
The last families of gates are diagonal in the computational basis, which are divided into 3 families: $D_2$, $D_3$ and $D_n$. Here, $D_2$ gates are applied to pairs of qubits, $D_3$ gates are applied to 3 qubits, and $D_n$ gates are applied to all the qubits. Diagonal circuits cannot perform universal computation but they are not always classically simulatable \cite{diagonals}. For more details about these families of quantum reservoirs see Appendix \ref{sect:QR}.

\subsubsection{Quantum machine learning task}
\label{sec:MLtask}
In order to assess the performance of the different quantum reservoirs, a machine learning task needs to be defined. In this work, the task consists of predicting the first excited electronic energy $E_1$ 
using only the associated ground state $\ket{\psi_0}_{R}$ with energy $E_0$ for the LiH and H$_2$O molecules. 
The ground states $\ket{\psi_0}_{R}$ for such Hamiltonians are calculated by exact 
diagonalization for different configuration ranges:
$R_{\text{LiH}} \in [0.5, 3.5]$ a.u., $R_{\text{OH}} \in [0.5,1.5]$ a.u.,
and $\phi_{\text{HOH}}=104.45^\circ$. The details of the ground state calculation are given in Ref.~\cite{CITA}. 
The ground state of the molecules is described using $n=8$ qubits for LiH and $n=10$ qubits for H$_2$O.
The datasets $\{\ket{\psi_0}_{R}, E_1(R)\}_R$ is split into training and test sets, 
where the test set contains the 30\% of the data $R_{\text{LiH}} \in [1.1, 2.0]$ a.u.\ and
$R_{\text{OH}} \in [1.05, 1.35]$ a.u.\,
and it is chosen so that the reservoir has to extrapolate to unseen data. 
The performance of the seven families of quantum reservoirs will be compared to the Krylov evolution.

\subsection{The Krylov approach to complexity}
 \label{sec:krylov}

The Krylov method is a numerical technique originally designed to approximate the action of a matrix 
or operator on a vector, without explicitly calculating the full matrix elements 
\cite{Hochbruck1997, Parlett1998}. 
It is particularly useful when dealing with large matrices or operators, which are common in quantum mechanics,
especially in the context of solving the time-dependent Schrödinger equation, 
Heisenberg evolution of operators, or eigenvalue problems \cite{park1986unitary,saad1992analysis,stewart1996error,hochbruck1997krylov,expokit,moler2003nineteen,Jawecki:2020cc,Ruffinelli2022}. 
Notably, it has been recently used to determine the complexity of quantum evolution for 
both operators and states \cite{PhysRevX.9.041017,rabinovici2021operator,balasubramanian2022quantum}. 
In this work, we focus on the second case, i.e.,~states.

Let us consider a state $\kp$ in a complex Hilbert space $\HC$ with dimensionality $N$, 
i.e.,~$\HC = \Cbb^N$. 
This state evolves under the influence of a time-independent Hamiltonian denoted by $H \in \rm End(\HC)$. We introduce the $D$-dimensional Krylov subspace, denoted as $\KC_D$,
\begin{equation}\label{eq:kry_subs}
    \KC_D = \text{span} \{ \kp , H \kp , \ldots, H^{D-1} \kp \}.
\end{equation}
To ensure generality, we assume that the state $\kp$ and the Hamiltonian $H$ do not share any symmetries,
meaning $\KC_N = \HC$. 
When there are shared symmetries, the time evolution is constrained within their respective subspaces. Consequently, the problem should be redefined to operate solely within that subspace.  

The Krylov method endeavors to approximate the time-evolved state $\kpt$ using the most optimal 
element from the set $\KC_D$.
To accomplish this, we create a set of orthonormal basis vectors, denoted as 
$B_D = \{ \ket{v_0} \equiv \kp, \ldots, \ket{v_{D-1}} \}$,  for the Krylov subspace. 
Usually, the basis is created using Lanczos's algorithm, which is a modified version of the 
Gram-Schmidt procedure. 
Lanczos's algorithm leverages the fact that orthonormalization is only required for the last two vectors 
in the basis. 
In order for the Krylov subspace to fully encompass the entire evolution at all times, it must span the 
entire Hilbert space, meaning that $D = N$, where $D$ represents the dimension of the Hilbert space. 
By following this approach, the Hamiltonian is transformed into its tridiagonal form,
\begin{equation}
    H \ket{v_k} = a_n \ket{v_k} + b_{k+1} \ket{v_{k+1}} + b_{k} \ket{v_{k-1}},
    \label{eq:H_tridiagonal}
\end{equation}
where $a_k$ and $b_k$ are values that define the Lanczos sequences.
This observation leads us to interpret the system on the Krylov basis, as a 1-D non-interacting 
tight-binding model. 
An initial state localized at one end of this effective chain evolves under the influence of the 
onsite potentials $a_k$ and hopping amplitudes $b_k$ at the nth site—causing the excitation to propagate 
and populate the rest of the lattice.

By expressing the evolved state $\ket{\psi(t)}$ in this Krylov basis, we have,
\begin{equation}
    \ket{\psi(t)} = \sum_{k=0}^{D-1} \psi_k(t) \ket{v_k}
    \label{eq:psit_krylov}
\end{equation}

The coefficients $\psi_k(t)$ can be obtained by solving the Schrödinger equation,
\begin{equation}
    i\ \dot{\psi}_k(t) = a_k \psi_k(t) + b_{k}\psi_{k-1}(t) +b_{k+1}\psi_{k+1}(t)
    \label{eq:psit_tightbinding}
\end{equation}


From this formalism, the concept of Krylov complexity denoted by $C_{\mathcal{K}}(t)$ arises,
which represents the average position,
\begin{equation}
    C_{\mathcal{K}}(t) = \sum_{k=0}^{D-1} k | \psi_k(t) |^2 .
    \label{eq:K_complexity}
\end{equation}
This complexity measure can be understood as the average dimension of the Krylov subspace needed to 
accurately represent the evolution of the initial state over time.

In recent studies \cite{Rabinovici_2022_integrability,camargo2023spectral,hashimoto2023krylov}, 
the potential of using Krylov complexity as a reliable measure to assess the transition from integrability to chaos has been explored. 
This investigation has been conducted in various systems, including those with and without semiclassical limits.
The results indicate that both the complexity and variance of the Lanczos coefficients provide an excellent
description of this transition when considering Krylov evolution for states. 

\section{Conclusions} \label{sec:conclusions}
In the realm of the rapidly progressing field of developing quantum methods for NISQ devices, 
which is aimed at surpassing classical approaches using the currently available quantum computers, 
QRC algorithm has emerged as a standout approach. 
This algorithm harnesses the inherent properties and dynamics of a quantum system, 
referred to as the quantum reservoir, to discover hidden patterns within the input data 
crucial to perform optimal machine learning tasks. 
Indeed, the design of the quantum reservoir is of uttermost importance for the successful implementation 
of the algorithm. 
In this respect, it was recently proven that the complexity of the quantum reservoir, 
measured by the majorization criterion \cite{majorization}, is an excellent indicator of performance 
in machine learning tasks \cite{CITA, domingo2023taking}. 

Building upon this knowledge, the present work expands the understanding by providing a quantitative 
and precise method to assess the performance of quantum reservoirs, employing the Krylov approach to complexity.
To achieve this, we first had to develop a technique able to calculate the Krylov evolution for quantum 
unitary operators. 
Through this approach, we successfully reproduced the chaotic phase transition of two quantum systems, 
namely the longitudinal-transverse field Ising Hamiltonian and the standard map, validating our
ansatz of Eq.~(\ref{eq:H_eff}). 

Subsequently, we demonstrated that Krylov statistics, including the mean $K-$complexity at 
saturation $\overline{C_k}$ and its variance Var($\overline{C_k}$), along with the variance 
of the Lanczos coefficients, exhibit a strong correlation with the performance of quantum reservoirs.
Consequently, the Krylov evolution emerges as an efficient and precise tool for estimating the 
performance of QRC. 
These findings hold immense promise for designing optimal quantum algorithms involving random quantum systems,
facilitating successful implementations of quantum machine learning methods.

\section*{Acknowledgments}
The project that gave rise to these results received the support of a fellowship from 
``la Caixa'' Foundation (ID 100010434). 
The fellowship code is LCF/BQ/DR20/11790028.
This work has also been partially supported by the Spanish Ministry of Science, 
Innovation and Universities, Gobierno de Espa\~na, under Contract No.\ PID2021-122711NB-C21, and
CONICET (PIP 11220200100568CO), UBACyT (20020130100406BA) and ANPCyT (PICT-2016-1056).
This project has received funding from the European Union's Horizon 2020 research and innovation program under the Marie Sklodowska-Curie grant agreement No.~777822.

\section*{Data availability}
The datasets generated and analysed during the current study are available in the GitHub repository, \href{https://github.com/laiadc/Krylov\_QR/}
{https://github.com/laiadc/Krylov\_QR/}.

\section*{Code availability}
\label{sect:code}
The underlying code for this study is available in the GitHub repository and can be accessed 
via this link \href{https://github.com/laiadc/Krylov\_QR/}
{https://github.com/laiadc/Krylov\_QR/}.

\section*{Author contribution statement}
All authors developed the idea and the theory. LD performed the calculations and analyzed the data. All authors contributed to the discussions and interpretations of the results and wrote the manuscript. 

\section*{Competing financial interest statement}
The authors declare to have no competing financial and non-financial interests.

\bibliography{bibliography}

\begin{thebibliography}{66}%
\makeatletter
\providecommand \@ifxundefined [1]{%
 \@ifx{#1\undefined}
}%
\providecommand \@ifnum [1]{%
 \ifnum #1\expandafter \@firstoftwo
 \else \expandafter \@secondoftwo
 \fi
}%
\providecommand \@ifx [1]{%
 \ifx #1\expandafter \@firstoftwo
 \else \expandafter \@secondoftwo
 \fi
}%
\providecommand \natexlab [1]{#1}%
\providecommand \enquote  [1]{``#1''}%
\providecommand \bibnamefont  [1]{#1}%
\providecommand \bibfnamefont [1]{#1}%
\providecommand \citenamefont [1]{#1}%
\providecommand \href@noop [0]{\@secondoftwo}%
\providecommand \href [0]{\begingroup \@sanitize@url \@href}%
\providecommand \@href[1]{\@@startlink{#1}\@@href}%
\providecommand \@@href[1]{\endgroup#1\@@endlink}%
\providecommand \@sanitize@url [0]{\catcode `\\12\catcode `\$12\catcode
  `\&12\catcode `\#12\catcode `\^12\catcode `\_12\catcode `\%12\relax}%
\providecommand \@@startlink[1]{}%
\providecommand \@@endlink[0]{}%
\providecommand \url  [0]{\begingroup\@sanitize@url \@url }%
\providecommand \@url [1]{\endgroup\@href {#1}{\urlprefix }}%
\providecommand \urlprefix  [0]{URL }%
\providecommand \Eprint [0]{\href }%
\providecommand \doibase [0]{http://dx.doi.org/}%
\providecommand \selectlanguage [0]{\@gobble}%
\providecommand \bibinfo  [0]{\@secondoftwo}%
\providecommand \bibfield  [0]{\@secondoftwo}%
\providecommand \translation [1]{[#1]}%
\providecommand \BibitemOpen [0]{}%
\providecommand \bibitemStop [0]{}%
\providecommand \bibitemNoStop [0]{.\EOS\space}%
\providecommand \EOS [0]{\spacefactor3000\relax}%
\providecommand \BibitemShut  [1]{\csname bibitem#1\endcsname}%
\let\auto@bib@innerbib\@empty
\bibitem [{\citenamefont {Shor}(1997)}]{Shor}%
  \BibitemOpen
  \bibfield  {author} {\bibinfo {author} {\bibfnamefont {P.~W.}\ \bibnamefont
  {Shor}},\ }\href@noop {} {\bibfield  {journal} {\bibinfo  {journal} {SIAM J.
  Comput.}\ }\textbf {\bibinfo {volume} {26}},\ \bibinfo {pages} {1484–1509}
  (\bibinfo {year} {1997})}\BibitemShut {NoStop}%
\bibitem [{\citenamefont {Grover}(1996)}]{grover}%
  \BibitemOpen
  \bibfield  {author} {\bibinfo {author} {\bibfnamefont {L.~K.}\ \bibnamefont
  {Grover}},\ }in\ \href@noop {} {\emph {\bibinfo {booktitle} {Proceedings of
  the Twenty-Eighth Annual ACM Symposium on Theory of Computing}}}\ (\bibinfo
  {publisher} {Association for Computing Machinery},\ \bibinfo {year} {1996})\
  p.\ \bibinfo {pages} {212–219}\BibitemShut {NoStop}%
\bibitem [{\citenamefont {Bharti}\ \emph {et~al.}(2022)\citenamefont {Bharti},
  \citenamefont {Cervera-Lierta}, \citenamefont {Kyaw}, \citenamefont {Haug},
  \citenamefont {Alperin-Lea}, \citenamefont {Anand}, \citenamefont {Degroote},
  \citenamefont {Heimonen}, \citenamefont {Kottmann}, \citenamefont {Menke},
  \citenamefont {Mok}, \citenamefont {Sim}, \citenamefont {Kwek},\ and\
  \citenamefont {Aspuru-Guzik}}]{NISQ}%
  \BibitemOpen
  \bibfield  {author} {\bibinfo {author} {\bibfnamefont {K.}~\bibnamefont
  {Bharti}}, \bibinfo {author} {\bibfnamefont {A.}~\bibnamefont
  {Cervera-Lierta}}, \bibinfo {author} {\bibfnamefont {T.~H.}\ \bibnamefont
  {Kyaw}}, \bibinfo {author} {\bibfnamefont {T.}~\bibnamefont {Haug}}, \bibinfo
  {author} {\bibfnamefont {S.}~\bibnamefont {Alperin-Lea}}, \bibinfo {author}
  {\bibfnamefont {A.}~\bibnamefont {Anand}}, \bibinfo {author} {\bibfnamefont
  {M.}~\bibnamefont {Degroote}}, \bibinfo {author} {\bibfnamefont
  {H.}~\bibnamefont {Heimonen}}, \bibinfo {author} {\bibfnamefont {J.~S.}\
  \bibnamefont {Kottmann}}, \bibinfo {author} {\bibfnamefont {T.}~\bibnamefont
  {Menke}}, \bibinfo {author} {\bibfnamefont {W.-K.}\ \bibnamefont {Mok}},
  \bibinfo {author} {\bibfnamefont {S.}~\bibnamefont {Sim}}, \bibinfo {author}
  {\bibfnamefont {L.~C.}\ \bibnamefont {Kwek}}, \ and\ \bibinfo {author}
  {\bibfnamefont {A.}~\bibnamefont {Aspuru-Guzik}},\ }\href@noop {} {\bibfield
  {journal} {\bibinfo  {journal} {Rev. Mod. Phys.}\ }\textbf {\bibinfo {volume}
  {94}},\ \bibinfo {pages} {015004} (\bibinfo {year} {2022})}\BibitemShut
  {NoStop}%
\bibitem [{\citenamefont {Preskill}(2018)}]{Preskill18}%
  \BibitemOpen
  \bibfield  {author} {\bibinfo {author} {\bibfnamefont {J.}~\bibnamefont
  {Preskill}},\ }\href@noop {} {\bibfield  {journal} {\bibinfo  {journal}
  {Quantum}\ }\textbf {\bibinfo {volume} {2}},\ \bibinfo {pages} {79} (\bibinfo
  {year} {2018})}\BibitemShut {NoStop}%
\bibitem [{\citenamefont {Fujii}\ and\ \citenamefont
  {Nakajima}(2021)}]{Fujii2021}%
  \BibitemOpen
  \bibfield  {author} {\bibinfo {author} {\bibfnamefont {K.}~\bibnamefont
  {Fujii}}\ and\ \bibinfo {author} {\bibfnamefont {K.}~\bibnamefont
  {Nakajima}},\ }\enquote {\bibinfo {title} {Quantum reservoir computing: A
  reservoir approach toward quantum machine learning on near-term quantum
  devices},}\ in\ \href@noop {} {\emph {\bibinfo {booktitle} {Reservoir
  Computing: Theory, Physical Implementations, and Applications}}},\ \bibinfo
  {editor} {edited by\ \bibinfo {editor} {\bibfnamefont {K.}~\bibnamefont
  {Nakajima}}\ and\ \bibinfo {editor} {\bibfnamefont {I.}~\bibnamefont
  {Fischer}}}\ (\bibinfo  {publisher} {Springer Singapore},\ \bibinfo {address}
  {Singapore},\ \bibinfo {year} {2021})\ pp.\ \bibinfo {pages}
  {423--450}\BibitemShut {NoStop}%
\bibitem [{\citenamefont {Ghosh}\ \emph {et~al.}(2019)\citenamefont {Ghosh},
  \citenamefont {Opala}, \citenamefont {Matuszewski}, \citenamefont {Paterek},\
  and\ \citenamefont {Liew}}]{QRC2}%
  \BibitemOpen
  \bibfield  {author} {\bibinfo {author} {\bibfnamefont {S.}~\bibnamefont
  {Ghosh}}, \bibinfo {author} {\bibfnamefont {A.}~\bibnamefont {Opala}},
  \bibinfo {author} {\bibfnamefont {M.}~\bibnamefont {Matuszewski}}, \bibinfo
  {author} {\bibfnamefont {T.}~\bibnamefont {Paterek}}, \ and\ \bibinfo
  {author} {\bibfnamefont {T.~C.~H.}\ \bibnamefont {Liew}},\ }\href@noop {}
  {\bibfield  {journal} {\bibinfo  {journal} {npj Quantum Inf.}\ }\textbf
  {\bibinfo {volume} {5}},\ \bibinfo {pages} {35} (\bibinfo {year}
  {2019})}\BibitemShut {NoStop}%
\bibitem [{\citenamefont {Mart\'{\i}nez-Pe\~na}\ \emph
  {et~al.}(2021)\citenamefont {Mart\'{\i}nez-Pe\~na}, \citenamefont {Giorgi},
  \citenamefont {Nokkala}, \citenamefont {Soriano},\ and\ \citenamefont
  {Zambrini}}]{DynamicalIsing}%
  \BibitemOpen
  \bibfield  {author} {\bibinfo {author} {\bibfnamefont {R.}~\bibnamefont
  {Mart\'{\i}nez-Pe\~na}}, \bibinfo {author} {\bibfnamefont {G.~L.}\
  \bibnamefont {Giorgi}}, \bibinfo {author} {\bibfnamefont {J.}~\bibnamefont
  {Nokkala}}, \bibinfo {author} {\bibfnamefont {M.~C.}\ \bibnamefont
  {Soriano}}, \ and\ \bibinfo {author} {\bibfnamefont {R.}~\bibnamefont
  {Zambrini}},\ }\href {\doibase 10.1103/PhysRevLett.127.100502} {\bibfield
  {journal} {\bibinfo  {journal} {Phys. Rev. Lett.}\ }\textbf {\bibinfo
  {volume} {127}},\ \bibinfo {pages} {100502} (\bibinfo {year}
  {2021})}\BibitemShut {NoStop}%
\bibitem [{\citenamefont {Kutvonen}\ \emph
  {et~al.}(2020{\natexlab{a}})\citenamefont {Kutvonen}, \citenamefont {Fujii},\
  and\ \citenamefont {Sagawa}}]{OptQRC}%
  \BibitemOpen
  \bibfield  {author} {\bibinfo {author} {\bibfnamefont {A.}~\bibnamefont
  {Kutvonen}}, \bibinfo {author} {\bibfnamefont {K.}~\bibnamefont {Fujii}}, \
  and\ \bibinfo {author} {\bibfnamefont {T.}~\bibnamefont {Sagawa}},\
  }\href@noop {} {\bibfield  {journal} {\bibinfo  {journal} {Sci. Rep.}\
  }\textbf {\bibinfo {volume} {10}},\ \bibinfo {pages} {14687} (\bibinfo {year}
  {2020}{\natexlab{a}})}\BibitemShut {NoStop}%
\bibitem [{\citenamefont {Domingo}\ \emph {et~al.}(2022)\citenamefont
  {Domingo}, \citenamefont {Carlo},\ and\ \citenamefont {Borondo}}]{CITA}%
  \BibitemOpen
  \bibfield  {author} {\bibinfo {author} {\bibfnamefont {L.}~\bibnamefont
  {Domingo}}, \bibinfo {author} {\bibfnamefont {G.~G.}\ \bibnamefont {Carlo}},
  \ and\ \bibinfo {author} {\bibfnamefont {F.}~\bibnamefont {Borondo}},\
  }\href@noop {} {\bibfield  {journal} {\bibinfo  {journal} {Phys. Rev. E}\
  }\textbf {\bibinfo {volume} {106}},\ \bibinfo {pages} {L043301} (\bibinfo
  {year} {2022})}\BibitemShut {NoStop}%
\bibitem [{\citenamefont {Domingo}\ \emph {et~al.}(2023)\citenamefont
  {Domingo}, \citenamefont {Carlo},\ and\ \citenamefont
  {Borondo}}]{domingo2023taking}%
  \BibitemOpen
  \bibfield  {author} {\bibinfo {author} {\bibfnamefont {L.}~\bibnamefont
  {Domingo}}, \bibinfo {author} {\bibfnamefont {G.}~\bibnamefont {Carlo}}, \
  and\ \bibinfo {author} {\bibfnamefont {F.}~\bibnamefont {Borondo}},\
  }\href@noop {} {\bibfield  {journal} {\bibinfo  {journal} {Sci. Rep.}\
  }\textbf {\bibinfo {volume} {13}},\ \bibinfo {pages} {8790} (\bibinfo {year}
  {2023})}\BibitemShut {NoStop}%
\bibitem [{\citenamefont {Jaeger}(2001)}]{ESN}%
  \BibitemOpen
  \bibfield  {author} {\bibinfo {author} {\bibfnamefont {H.}~\bibnamefont
  {Jaeger}},\ }\href
  {http://www.faculty.jacobs-university.de/hjaeger/pubs/EchoStatesTechRep.pdf}
  {\bibfield  {journal} {\bibinfo  {journal} {German National Research Center
  for Information Technology GMD Technical Report}\ }\textbf {\bibinfo {volume}
  {148}} (\bibinfo {year} {2001})}\BibitemShut {NoStop}%
\bibitem [{\citenamefont {Lu}\ \emph {et~al.}(2022)\citenamefont {Lu},
  \citenamefont {Krifyuk}, \citenamefont {Zhang}, \citenamefont {Gao},
  \citenamefont {Lu},\ and\ \citenamefont {Sun}}]{imageQRC}%
  \BibitemOpen
  \bibfield  {author} {\bibinfo {author} {\bibfnamefont {D.}~\bibnamefont
  {Lu}}, \bibinfo {author} {\bibfnamefont {D.}~\bibnamefont {Krifyuk}},
  \bibinfo {author} {\bibfnamefont {Y.}~\bibnamefont {Zhang}}, \bibinfo
  {author} {\bibfnamefont {X.}~\bibnamefont {Gao}}, \bibinfo {author}
  {\bibfnamefont {K.}~\bibnamefont {Lu}}, \ and\ \bibinfo {author}
  {\bibfnamefont {J.}~\bibnamefont {Sun}},\ }\href {\doibase
  10.1103/PRXQuantum.3.010301} {\bibfield  {journal} {\bibinfo  {journal} {PRX
  Quantum}\ }\textbf {\bibinfo {volume} {3}},\ \bibinfo {pages} {010301}
  (\bibinfo {year} {2022})}\BibitemShut {NoStop}%
\bibitem [{\citenamefont {Kawai}\ and\ \citenamefont
  {Nakagawa}(2020)}]{quantumchemQRC}%
  \BibitemOpen
  \bibfield  {author} {\bibinfo {author} {\bibfnamefont {H.}~\bibnamefont
  {Kawai}}\ and\ \bibinfo {author} {\bibfnamefont {Y.}~\bibnamefont
  {Nakagawa}},\ }\href@noop {} {\bibfield  {journal} {\bibinfo  {journal}
  {Mach. Learn.: Sci. Technol.}\ }\textbf {\bibinfo {volume} {1}} (\bibinfo
  {year} {2020})}\BibitemShut {NoStop}%
\bibitem [{\citenamefont {Mujal}\ \emph {et~al.}(2023)\citenamefont {Mujal},
  \citenamefont {Martínez-Peña}, \citenamefont {Giorgi}, \citenamefont
  {Soriano},\ and\ \citenamefont {Zambrini}}]{zambrini_npj}%
  \BibitemOpen
  \bibfield  {author} {\bibinfo {author} {\bibfnamefont {P.}~\bibnamefont
  {Mujal}}, \bibinfo {author} {\bibfnamefont {R.}~\bibnamefont
  {Martínez-Peña}}, \bibinfo {author} {\bibfnamefont {G.~L.}\ \bibnamefont
  {Giorgi}}, \bibinfo {author} {\bibfnamefont {M.~C.}\ \bibnamefont {Soriano}},
  \ and\ \bibinfo {author} {\bibfnamefont {R.}~\bibnamefont {Zambrini}},\
  }\href {\doibase 10.1038/s41534-023-00682-z} {\bibfield  {journal} {\bibinfo
  {journal} {{npj} Quantum Inf.}\ }\textbf {\bibinfo {volume} {9}},\ \bibinfo
  {pages} {16} (\bibinfo {year} {2023})}\BibitemShut {NoStop}%
\bibitem [{\citenamefont {Nokkala}\ \emph {et~al.}(2021)\citenamefont
  {Nokkala}, \citenamefont {Martínez-Peña}, \citenamefont {Giorgi},
  \citenamefont {Parigi}, \citenamefont {Soriano},\ and\ \citenamefont
  {Zambrini}}]{zambrini_com_phys}%
  \BibitemOpen
  \bibfield  {author} {\bibinfo {author} {\bibfnamefont {J.}~\bibnamefont
  {Nokkala}}, \bibinfo {author} {\bibfnamefont {R.}~\bibnamefont
  {Martínez-Peña}}, \bibinfo {author} {\bibfnamefont {G.~L.}\ \bibnamefont
  {Giorgi}}, \bibinfo {author} {\bibfnamefont {V.}~\bibnamefont {Parigi}},
  \bibinfo {author} {\bibfnamefont {M.~C.}\ \bibnamefont {Soriano}}, \ and\
  \bibinfo {author} {\bibfnamefont {R.}~\bibnamefont {Zambrini}},\ }\href
  {\doibase 10.1038/s42005-021-00556-w} {\bibfield  {journal} {\bibinfo
  {journal} {Commun. Phys.}\ }\textbf {\bibinfo {volume} {4}},\ \bibinfo
  {pages} {53} (\bibinfo {year} {2021})}\BibitemShut {NoStop}%
\bibitem [{\citenamefont {Nakajima}\ \emph {et~al.}(2019)\citenamefont
  {Nakajima}, \citenamefont {Fujii}, \citenamefont {Negoro}, \citenamefont
  {Mitarai},\ and\ \citenamefont {Kitagawa}}]{time_series2}%
  \BibitemOpen
  \bibfield  {author} {\bibinfo {author} {\bibfnamefont {K.}~\bibnamefont
  {Nakajima}}, \bibinfo {author} {\bibfnamefont {K.}~\bibnamefont {Fujii}},
  \bibinfo {author} {\bibfnamefont {M.}~\bibnamefont {Negoro}}, \bibinfo
  {author} {\bibfnamefont {K.}~\bibnamefont {Mitarai}}, \ and\ \bibinfo
  {author} {\bibfnamefont {M.}~\bibnamefont {Kitagawa}},\ }\href {\doibase
  10.1103/PhysRevApplied.11.034021} {\bibfield  {journal} {\bibinfo  {journal}
  {Phys. Rev. Appl.}\ }\textbf {\bibinfo {volume} {11}},\ \bibinfo {pages}
  {034021} (\bibinfo {year} {2019})}\BibitemShut {NoStop}%
\bibitem [{\citenamefont {Kutvonen}\ \emph
  {et~al.}(2020{\natexlab{b}})\citenamefont {Kutvonen}, \citenamefont {Fujii},\
  and\ \citenamefont {Sagawa}}]{time_series3}%
  \BibitemOpen
  \bibfield  {author} {\bibinfo {author} {\bibfnamefont {A.}~\bibnamefont
  {Kutvonen}}, \bibinfo {author} {\bibfnamefont {K.}~\bibnamefont {Fujii}}, \
  and\ \bibinfo {author} {\bibfnamefont {T.}~\bibnamefont {Sagawa}},\ }\href
  {\doibase 10.1038/s41598-020-71673-9} {\bibfield  {journal} {\bibinfo
  {journal} {Sci. Rep.}\ }\textbf {\bibinfo {volume} {10}},\ \bibinfo {pages}
  {14687} (\bibinfo {year} {2020}{\natexlab{b}})}\BibitemShut {NoStop}%
\bibitem [{\citenamefont {Chen}\ \emph
  {et~al.}(2020{\natexlab{a}})\citenamefont {Chen}, \citenamefont {Nurdin},\
  and\ \citenamefont {Yamamoto}}]{time_Series4}%
  \BibitemOpen
  \bibfield  {author} {\bibinfo {author} {\bibfnamefont {J.}~\bibnamefont
  {Chen}}, \bibinfo {author} {\bibfnamefont {H.~I.}\ \bibnamefont {Nurdin}}, \
  and\ \bibinfo {author} {\bibfnamefont {N.}~\bibnamefont {Yamamoto}},\ }\href
  {\doibase 10.1103/PhysRevApplied.14.024065} {\bibfield  {journal} {\bibinfo
  {journal} {Phys. Rev. Appl.}\ }\textbf {\bibinfo {volume} {14}},\ \bibinfo
  {pages} {024065} (\bibinfo {year} {2020}{\natexlab{a}})}\BibitemShut
  {NoStop}%
\bibitem [{\citenamefont {Mart{\'\i}nez-Pe{\~n}a}\ \emph
  {et~al.}(2020)\citenamefont {Mart{\'\i}nez-Pe{\~n}a}, \citenamefont
  {Nokkala}, \citenamefont {Giorgi},\ and\ \citenamefont
  {Amato-Grill}}]{time_series5}%
  \BibitemOpen
  \bibfield  {author} {\bibinfo {author} {\bibfnamefont {R.}~\bibnamefont
  {Mart{\'\i}nez-Pe{\~n}a}}, \bibinfo {author} {\bibfnamefont {J.}~\bibnamefont
  {Nokkala}}, \bibinfo {author} {\bibfnamefont {G.~L.}\ \bibnamefont {Giorgi}},
  \ and\ \bibinfo {author} {\bibfnamefont {J.}~\bibnamefont {Amato-Grill}},\
  }\href {\doibase 10.1007/s12559-020-09772-y} {\bibfield  {journal} {\bibinfo
  {journal} {Cogn. Comput.}\ }\textbf {\bibinfo {volume} {12}},\ \bibinfo
  {pages} {1} (\bibinfo {year} {2020})}\BibitemShut {NoStop}%
\bibitem [{\citenamefont {Latorre}\ and\ \citenamefont
  {Mart\'{\i}n-Delgado}(2002)}]{majorization_original}%
  \BibitemOpen
  \bibfield  {author} {\bibinfo {author} {\bibfnamefont {J.~I.}\ \bibnamefont
  {Latorre}}\ and\ \bibinfo {author} {\bibfnamefont {M.~A.}\ \bibnamefont
  {Mart\'{\i}n-Delgado}},\ }\href {\doibase 10.1103/PhysRevA.66.022305}
  {\bibfield  {journal} {\bibinfo  {journal} {Phys. Rev. A}\ }\textbf {\bibinfo
  {volume} {66}},\ \bibinfo {pages} {022305} (\bibinfo {year}
  {2002})}\BibitemShut {NoStop}%
\bibitem [{\citenamefont {Vallejos}\ \emph
  {et~al.}(2021{\natexlab{a}})\citenamefont {Vallejos}, \citenamefont
  {de~Melo},\ and\ \citenamefont {Carlo}}]{majorization}%
  \BibitemOpen
  \bibfield  {author} {\bibinfo {author} {\bibfnamefont {R.}~\bibnamefont
  {Vallejos}}, \bibinfo {author} {\bibfnamefont {F.}~\bibnamefont {de~Melo}}, \
  and\ \bibinfo {author} {\bibfnamefont {G.~G.}\ \bibnamefont {Carlo}},\
  }\href@noop {} {\bibfield  {journal} {\bibinfo  {journal} {Phys. Rev. A}\
  }\textbf {\bibinfo {volume} {104}},\ \bibinfo {pages} {012602} (\bibinfo
  {year} {2021}{\natexlab{a}})}\BibitemShut {NoStop}%
\bibitem [{\citenamefont {Latorre}\ and\ \citenamefont
  {Mart{\'\i}n-Delgado}(2002)}]{latorre2002majorization}%
  \BibitemOpen
  \bibfield  {author} {\bibinfo {author} {\bibfnamefont {J.~I.}\ \bibnamefont
  {Latorre}}\ and\ \bibinfo {author} {\bibfnamefont {M.}~\bibnamefont
  {Mart{\'\i}n-Delgado}},\ }\href@noop {} {\bibfield  {journal} {\bibinfo
  {journal} {Phys. Rev. A}\ }\textbf {\bibinfo {volume} {66}},\ \bibinfo
  {pages} {022305} (\bibinfo {year} {2002})}\BibitemShut {NoStop}%
\bibitem [{\citenamefont {Vallejos}\ \emph
  {et~al.}(2021{\natexlab{b}})\citenamefont {Vallejos}, \citenamefont
  {De~Melo},\ and\ \citenamefont {Carlo}}]{vallejos2021principle}%
  \BibitemOpen
  \bibfield  {author} {\bibinfo {author} {\bibfnamefont {R.~O.}\ \bibnamefont
  {Vallejos}}, \bibinfo {author} {\bibfnamefont {F.}~\bibnamefont {De~Melo}}, \
  and\ \bibinfo {author} {\bibfnamefont {G.~G.}\ \bibnamefont {Carlo}},\
  }\href@noop {} {\bibfield  {journal} {\bibinfo  {journal} {Phys. Rev. A}\
  }\textbf {\bibinfo {volume} {104}},\ \bibinfo {pages} {012602} (\bibinfo
  {year} {2021}{\natexlab{b}})}\BibitemShut {NoStop}%
\bibitem [{\citenamefont {Parker}\ \emph
  {et~al.}(2019{\natexlab{a}})\citenamefont {Parker}, \citenamefont {Cao},
  \citenamefont {Avdoshkin}, \citenamefont {Scaffidi},\ and\ \citenamefont
  {Altman}}]{Parker2019}%
  \BibitemOpen
  \bibfield  {author} {\bibinfo {author} {\bibfnamefont {D.~E.}\ \bibnamefont
  {Parker}}, \bibinfo {author} {\bibfnamefont {X.}~\bibnamefont {Cao}},
  \bibinfo {author} {\bibfnamefont {A.}~\bibnamefont {Avdoshkin}}, \bibinfo
  {author} {\bibfnamefont {T.}~\bibnamefont {Scaffidi}}, \ and\ \bibinfo
  {author} {\bibfnamefont {E.}~\bibnamefont {Altman}},\ }\href {\doibase
  10.1103/physrevx.9.041017} {\bibfield  {journal} {\bibinfo  {journal} {Phys.
  Rev. X}\ }\textbf {\bibinfo {volume} {9}},\ \bibinfo {pages} {041017}
  (\bibinfo {year} {2019}{\natexlab{a}})}\BibitemShut {NoStop}%
\bibitem [{\citenamefont {Rabinovici}\ \emph
  {et~al.}(2021{\natexlab{a}})\citenamefont {Rabinovici}, \citenamefont
  {S{\'a}nchez-Garrido}, \citenamefont {Shir},\ and\ \citenamefont
  {Sonner}}]{rabinovici2021operator}%
  \BibitemOpen
  \bibfield  {author} {\bibinfo {author} {\bibfnamefont {E.}~\bibnamefont
  {Rabinovici}}, \bibinfo {author} {\bibfnamefont {A.}~\bibnamefont
  {S{\'a}nchez-Garrido}}, \bibinfo {author} {\bibfnamefont {R.}~\bibnamefont
  {Shir}}, \ and\ \bibinfo {author} {\bibfnamefont {J.}~\bibnamefont
  {Sonner}},\ }\href {\doibase 10.1007/JHEP06(2021)062} {\bibfield  {journal}
  {\bibinfo  {journal} {J. High Energy Phys.}\ }\textbf {\bibinfo {volume}
  {2021}},\ \bibinfo {pages} {62} (\bibinfo {year}
  {2021}{\natexlab{a}})}\BibitemShut {NoStop}%
\bibitem [{\citenamefont {Parker}\ \emph
  {et~al.}(2019{\natexlab{b}})\citenamefont {Parker}, \citenamefont {Cao},
  \citenamefont {Avdoshkin}, \citenamefont {Scaffidi},\ and\ \citenamefont
  {Altman}}]{PhysRevX.9.041017}%
  \BibitemOpen
  \bibfield  {author} {\bibinfo {author} {\bibfnamefont {D.~E.}\ \bibnamefont
  {Parker}}, \bibinfo {author} {\bibfnamefont {X.}~\bibnamefont {Cao}},
  \bibinfo {author} {\bibfnamefont {A.}~\bibnamefont {Avdoshkin}}, \bibinfo
  {author} {\bibfnamefont {T.}~\bibnamefont {Scaffidi}}, \ and\ \bibinfo
  {author} {\bibfnamefont {E.}~\bibnamefont {Altman}},\ }\href {\doibase
  10.1103/PhysRevX.9.041017} {\bibfield  {journal} {\bibinfo  {journal} {Phys.
  Rev. X}\ }\textbf {\bibinfo {volume} {9}},\ \bibinfo {pages} {041017}
  (\bibinfo {year} {2019}{\natexlab{b}})}\BibitemShut {NoStop}%
\bibitem [{\citenamefont {Balasubramanian}\ \emph {et~al.}(2022)\citenamefont
  {Balasubramanian}, \citenamefont {Caputa}, \citenamefont {Magan},\ and\
  \citenamefont {Wu}}]{balasubramanian2022quantum}%
  \BibitemOpen
  \bibfield  {author} {\bibinfo {author} {\bibfnamefont {V.}~\bibnamefont
  {Balasubramanian}}, \bibinfo {author} {\bibfnamefont {P.}~\bibnamefont
  {Caputa}}, \bibinfo {author} {\bibfnamefont {J.~M.}\ \bibnamefont {Magan}}, \
  and\ \bibinfo {author} {\bibfnamefont {Q.}~\bibnamefont {Wu}},\ }\href@noop
  {} {\bibfield  {journal} {\bibinfo  {journal} {Phys. Rev. D}\ }\textbf
  {\bibinfo {volume} {106}},\ \bibinfo {pages} {046007} (\bibinfo {year}
  {2022})}\BibitemShut {NoStop}%
\bibitem [{\citenamefont {Trigueros}\ and\ \citenamefont
  {Lin}(2022)}]{Ballar_Trigueros2022}%
  \BibitemOpen
  \bibfield  {author} {\bibinfo {author} {\bibfnamefont {F.~B.}\ \bibnamefont
  {Trigueros}}\ and\ \bibinfo {author} {\bibfnamefont {C.-J.}\ \bibnamefont
  {Lin}},\ }\href {\doibase 10.21468/scipostphys.13.2.037} {\bibfield
  {journal} {\bibinfo  {journal} {{SciPost} Phys.}\ }\textbf {\bibinfo {volume}
  {13}} (\bibinfo {year} {2022}),\ 10.21468/scipostphys.13.2.037}\BibitemShut
  {NoStop}%
\bibitem [{\citenamefont {Adhikari}\ \emph {et~al.}(2023)\citenamefont
  {Adhikari}, \citenamefont {Choudhury},\ and\ \citenamefont {Roy}}]{KryQFT}%
  \BibitemOpen
  \bibfield  {author} {\bibinfo {author} {\bibfnamefont {K.}~\bibnamefont
  {Adhikari}}, \bibinfo {author} {\bibfnamefont {S.}~\bibnamefont {Choudhury}},
  \ and\ \bibinfo {author} {\bibfnamefont {A.}~\bibnamefont {Roy}},\ }\href
  {\doibase 10.1016/j.nuclphysb.2023.116263} {\bibfield  {journal} {\bibinfo
  {journal} {Nucl. Phys. B}\ }\textbf {\bibinfo {volume} {993}},\ \bibinfo
  {pages} {116263} (\bibinfo {year} {2023})}\BibitemShut {NoStop}%
\bibitem [{\citenamefont {Shen}\ \emph {et~al.}(2023)\citenamefont {Shen},
  \citenamefont {Klymko}, \citenamefont {Sud}, \citenamefont {Williams-Young},
  \citenamefont {de~Jong},\ and\ \citenamefont {Tubman}}]{KryQIT}%
  \BibitemOpen
  \bibfield  {author} {\bibinfo {author} {\bibfnamefont {Y.}~\bibnamefont
  {Shen}}, \bibinfo {author} {\bibfnamefont {K.}~\bibnamefont {Klymko}},
  \bibinfo {author} {\bibfnamefont {J.}~\bibnamefont {Sud}}, \bibinfo {author}
  {\bibfnamefont {D.~B.}\ \bibnamefont {Williams-Young}}, \bibinfo {author}
  {\bibfnamefont {W.~A.}\ \bibnamefont {de~Jong}}, \ and\ \bibinfo {author}
  {\bibfnamefont {N.~M.}\ \bibnamefont {Tubman}},\ }\href {\doibase
  10.22331/q-2023-07-25-1066} {\bibfield  {journal} {\bibinfo  {journal}
  {Quantum}\ }\textbf {\bibinfo {volume} {7}},\ \bibinfo {pages} {1066}
  (\bibinfo {year} {2023})}\BibitemShut {NoStop}%
\bibitem [{\citenamefont {Hochbruck}\ and\ \citenamefont
  {Lubich}(1997{\natexlab{a}})}]{Hochbruck1997}%
  \BibitemOpen
  \bibfield  {author} {\bibinfo {author} {\bibfnamefont {M.}~\bibnamefont
  {Hochbruck}}\ and\ \bibinfo {author} {\bibfnamefont {C.}~\bibnamefont
  {Lubich}},\ }\href {http://www.jstor.org/stable/2952023} {\bibfield
  {journal} {\bibinfo  {journal} {SIAM J. Numer. Anal.}\ }\textbf {\bibinfo
  {volume} {34}},\ \bibinfo {pages} {1911} (\bibinfo {year}
  {1997}{\natexlab{a}})}\BibitemShut {NoStop}%
\bibitem [{\citenamefont {Parlett}(1987)}]{Parlett1998}%
  \BibitemOpen
  \bibfield  {author} {\bibinfo {author} {\bibfnamefont {B.~N.}\ \bibnamefont
  {Parlett}},\ }\href@noop {} {\emph {\bibinfo {title} {The Symmetric
  Eigenvalue Problem (Classics in Applied Mathematics, Series Number 20)}}}\
  (\bibinfo  {publisher} {Society for Industrial and Applied Mathematics},\
  \bibinfo {year} {1987})\BibitemShut {NoStop}%
\bibitem [{\citenamefont {Ruffinelli}\ \emph {et~al.}(2022)\citenamefont
  {Ruffinelli}, \citenamefont {Fortes}, \citenamefont {Wisniacki},\ and\
  \citenamefont {Larocca}}]{Ruffinelli2022}%
  \BibitemOpen
  \bibfield  {author} {\bibinfo {author} {\bibfnamefont {J.~M.}\ \bibnamefont
  {Ruffinelli}}, \bibinfo {author} {\bibfnamefont {E.~M.}\ \bibnamefont
  {Fortes}}, \bibinfo {author} {\bibfnamefont {D.~A.}\ \bibnamefont
  {Wisniacki}}, \ and\ \bibinfo {author} {\bibfnamefont {M.}~\bibnamefont
  {Larocca}},\ }\href {\doibase 10.1103/physreva.106.042423} {\bibfield
  {journal} {\bibinfo  {journal} {Phys. Rev. A}\ }\textbf {\bibinfo {volume}
  {106}} (\bibinfo {year} {2022}),\ 10.1103/physreva.106.042423}\BibitemShut
  {NoStop}%
\bibitem [{\citenamefont {Dymarsky}\ and\ \citenamefont
  {Gorsky}(2020)}]{Dymarsky2020}%
  \BibitemOpen
  \bibfield  {author} {\bibinfo {author} {\bibfnamefont {A.}~\bibnamefont
  {Dymarsky}}\ and\ \bibinfo {author} {\bibfnamefont {A.}~\bibnamefont
  {Gorsky}},\ }\href {\doibase 10.1103/physrevb.102.085137} {\bibfield
  {journal} {\bibinfo  {journal} {Phys. Rev. B}\ }\textbf {\bibinfo {volume}
  {102}} (\bibinfo {year} {2020}),\ 10.1103/physrevb.102.085137}\BibitemShut
  {NoStop}%
\bibitem [{\citenamefont {Cao}(2021)}]{Cao2021}%
  \BibitemOpen
  \bibfield  {author} {\bibinfo {author} {\bibfnamefont {X.}~\bibnamefont
  {Cao}},\ }\href {\doibase 10.1088/1751-8121/abe77c} {\bibfield  {journal}
  {\bibinfo  {journal} {J. Phys. A}\ }\textbf {\bibinfo {volume} {54}},\
  \bibinfo {pages} {144001} (\bibinfo {year} {2021})}\BibitemShut {NoStop}%
\bibitem [{\citenamefont {Bhattacharjee}\ \emph
  {et~al.}(2022{\natexlab{a}})\citenamefont {Bhattacharjee}, \citenamefont
  {Nandy},\ and\ \citenamefont {Pathak}}]{Bhattacharjee2022_2}%
  \BibitemOpen
  \bibfield  {author} {\bibinfo {author} {\bibfnamefont {B.}~\bibnamefont
  {Bhattacharjee}}, \bibinfo {author} {\bibfnamefont {P.}~\bibnamefont
  {Nandy}}, \ and\ \bibinfo {author} {\bibfnamefont {T.}~\bibnamefont
  {Pathak}},\ }\href@noop {} {\bibfield  {journal} {\bibinfo  {journal}
  {arXiv}\ } (\bibinfo {year} {2022}{\natexlab{a}})},\ \Eprint
  {http://arxiv.org/abs/2210.02474v2} {2210.02474v2} \BibitemShut {NoStop}%
\bibitem [{\citenamefont {Bhattacharya}\ \emph {et~al.}(2022)\citenamefont
  {Bhattacharya}, \citenamefont {Nandy}, \citenamefont {Nath},\ and\
  \citenamefont {Sahu}}]{Bhattacharya2022}%
  \BibitemOpen
  \bibfield  {author} {\bibinfo {author} {\bibfnamefont {A.}~\bibnamefont
  {Bhattacharya}}, \bibinfo {author} {\bibfnamefont {P.}~\bibnamefont {Nandy}},
  \bibinfo {author} {\bibfnamefont {P.~P.}\ \bibnamefont {Nath}}, \ and\
  \bibinfo {author} {\bibfnamefont {H.}~\bibnamefont {Sahu}},\ }\href {\doibase
  10.1007/jhep12(2022)081} {\bibfield  {journal} {\bibinfo  {journal} {J. High
  Energy Phys.}\ }\textbf {\bibinfo {volume} {2022}} (\bibinfo {year} {2022}),\
  10.1007/jhep12(2022)081}\BibitemShut {NoStop}%
\bibitem [{\citenamefont {Bhattacharjee}\ \emph
  {et~al.}(2022{\natexlab{b}})\citenamefont {Bhattacharjee}, \citenamefont
  {Cao}, \citenamefont {Nandy},\ and\ \citenamefont
  {Pathak}}]{Bhattacharjee2022}%
  \BibitemOpen
  \bibfield  {author} {\bibinfo {author} {\bibfnamefont {B.}~\bibnamefont
  {Bhattacharjee}}, \bibinfo {author} {\bibfnamefont {X.}~\bibnamefont {Cao}},
  \bibinfo {author} {\bibfnamefont {P.}~\bibnamefont {Nandy}}, \ and\ \bibinfo
  {author} {\bibfnamefont {T.}~\bibnamefont {Pathak}},\ }\href {\doibase
  10.1007/jhep05(2022)174} {\bibfield  {journal} {\bibinfo  {journal} {J. High
  Energy Phys.}\ }\textbf {\bibinfo {volume} {2022}} (\bibinfo {year}
  {2022}{\natexlab{b}}),\ 10.1007/jhep05(2022)174}\BibitemShut {NoStop}%
\bibitem [{\citenamefont {Espa\~nol}\ and\ \citenamefont
  {Wisniacki}(2023)}]{PhysRevE.107.024217}%
  \BibitemOpen
  \bibfield  {author} {\bibinfo {author} {\bibfnamefont {B.~L.}\ \bibnamefont
  {Espa\~nol}}\ and\ \bibinfo {author} {\bibfnamefont {D.~A.}\ \bibnamefont
  {Wisniacki}},\ }\href {\doibase 10.1103/PhysRevE.107.024217} {\bibfield
  {journal} {\bibinfo  {journal} {Phys. Rev. E}\ }\textbf {\bibinfo {volume}
  {107}},\ \bibinfo {pages} {024217} (\bibinfo {year} {2023})}\BibitemShut
  {NoStop}%
\bibitem [{\citenamefont {Barb{\'o}n}\ \emph {et~al.}(2019)\citenamefont
  {Barb{\'o}n}, \citenamefont {Rabinovici}, \citenamefont {Shir},\ and\
  \citenamefont {Sinha}}]{barbon2019evolution}%
  \BibitemOpen
  \bibfield  {author} {\bibinfo {author} {\bibfnamefont {J.}~\bibnamefont
  {Barb{\'o}n}}, \bibinfo {author} {\bibfnamefont {E.}~\bibnamefont
  {Rabinovici}}, \bibinfo {author} {\bibfnamefont {R.}~\bibnamefont {Shir}}, \
  and\ \bibinfo {author} {\bibfnamefont {R.}~\bibnamefont {Sinha}},\
  }\href@noop {} {\bibfield  {journal} {\bibinfo  {journal} {J. High Energy
  Phys.}\ }\textbf {\bibinfo {volume} {2019}},\ \bibinfo {pages} {1} (\bibinfo
  {year} {2019})}\BibitemShut {NoStop}%
\bibitem [{\citenamefont {Rabinovici}\ \emph
  {et~al.}(2022{\natexlab{a}})\citenamefont {Rabinovici}, \citenamefont
  {S{\'{a}}nchez-Garrido}, \citenamefont {Shir},\ and\ \citenamefont
  {Sonner}}]{Rabinovici_2022_localization}%
  \BibitemOpen
  \bibfield  {author} {\bibinfo {author} {\bibfnamefont {E.}~\bibnamefont
  {Rabinovici}}, \bibinfo {author} {\bibfnamefont {A.}~\bibnamefont
  {S{\'{a}}nchez-Garrido}}, \bibinfo {author} {\bibfnamefont {R.}~\bibnamefont
  {Shir}}, \ and\ \bibinfo {author} {\bibfnamefont {J.}~\bibnamefont
  {Sonner}},\ }\href {\doibase 10.1007/jhep03(2022)211} {\bibfield  {journal}
  {\bibinfo  {journal} {J. High Energy Phys.}\ }\textbf {\bibinfo {volume}
  {2022}} (\bibinfo {year} {2022}{\natexlab{a}}),\
  10.1007/jhep03(2022)211}\BibitemShut {NoStop}%
\bibitem [{\citenamefont {Rabinovici}\ \emph
  {et~al.}(2021{\natexlab{b}})\citenamefont {Rabinovici}, \citenamefont
  {S{\'{a}}nchez-Garrido}, \citenamefont {Shir},\ and\ \citenamefont
  {Sonner}}]{Rabinovici2021}%
  \BibitemOpen
  \bibfield  {author} {\bibinfo {author} {\bibfnamefont {E.}~\bibnamefont
  {Rabinovici}}, \bibinfo {author} {\bibfnamefont {A.}~\bibnamefont
  {S{\'{a}}nchez-Garrido}}, \bibinfo {author} {\bibfnamefont {R.}~\bibnamefont
  {Shir}}, \ and\ \bibinfo {author} {\bibfnamefont {J.}~\bibnamefont
  {Sonner}},\ }\href {\doibase 10.1007/jhep06(2021)062} {\bibfield  {journal}
  {\bibinfo  {journal} {J. High Energy Phys.}\ }\textbf {\bibinfo {volume}
  {2021}} (\bibinfo {year} {2021}{\natexlab{b}}),\
  10.1007/jhep06(2021)062}\BibitemShut {NoStop}%
\bibitem [{\citenamefont {Rabinovici}\ \emph
  {et~al.}(2022{\natexlab{b}})\citenamefont {Rabinovici}, \citenamefont
  {S{\'{a}}nchez-Garrido}, \citenamefont {Shir},\ and\ \citenamefont
  {Sonner}}]{Rabinovici_sup2022}%
  \BibitemOpen
  \bibfield  {author} {\bibinfo {author} {\bibfnamefont {E.}~\bibnamefont
  {Rabinovici}}, \bibinfo {author} {\bibfnamefont {A.}~\bibnamefont
  {S{\'{a}}nchez-Garrido}}, \bibinfo {author} {\bibfnamefont {R.}~\bibnamefont
  {Shir}}, \ and\ \bibinfo {author} {\bibfnamefont {J.}~\bibnamefont
  {Sonner}},\ }\href {\doibase 10.1007/jhep03(2022)211} {\bibfield  {journal}
  {\bibinfo  {journal} {J. High Energy Phys.}\ }\textbf {\bibinfo {volume}
  {2022}} (\bibinfo {year} {2022}{\natexlab{b}}),\
  10.1007/jhep03(2022)211}\BibitemShut {NoStop}%
\bibitem [{\citenamefont {Rabinovici}\ \emph
  {et~al.}(2022{\natexlab{c}})\citenamefont {Rabinovici}, \citenamefont
  {S{\'{a}}nchez-Garrido}, \citenamefont {Shir},\ and\ \citenamefont
  {Sonner}}]{Rabinovici_2022_integrability}%
  \BibitemOpen
  \bibfield  {author} {\bibinfo {author} {\bibfnamefont {E.}~\bibnamefont
  {Rabinovici}}, \bibinfo {author} {\bibfnamefont {A.}~\bibnamefont
  {S{\'{a}}nchez-Garrido}}, \bibinfo {author} {\bibfnamefont {R.}~\bibnamefont
  {Shir}}, \ and\ \bibinfo {author} {\bibfnamefont {J.}~\bibnamefont
  {Sonner}},\ }\href {\doibase 10.1007/jhep07(2022)151} {\bibfield  {journal}
  {\bibinfo  {journal} {J. High Energy Phys.}\ }\textbf {\bibinfo {volume}
  {2022}} (\bibinfo {year} {2022}{\natexlab{c}}),\
  10.1007/jhep07(2022)151}\BibitemShut {NoStop}%
\bibitem [{\citenamefont {Camargo}\ \emph {et~al.}(2023)\citenamefont
  {Camargo}, \citenamefont {Jahnke}, \citenamefont {Jeong}, \citenamefont
  {Kim},\ and\ \citenamefont {Nishida}}]{camargo2023spectral}%
  \BibitemOpen
  \bibfield  {author} {\bibinfo {author} {\bibfnamefont {H.~A.}\ \bibnamefont
  {Camargo}}, \bibinfo {author} {\bibfnamefont {V.}~\bibnamefont {Jahnke}},
  \bibinfo {author} {\bibfnamefont {H.-S.}\ \bibnamefont {Jeong}}, \bibinfo
  {author} {\bibfnamefont {K.-Y.}\ \bibnamefont {Kim}}, \ and\ \bibinfo
  {author} {\bibfnamefont {M.}~\bibnamefont {Nishida}},\ }\href@noop {}
  {\bibfield  {journal} {\bibinfo  {journal} {arXiv preprint arXiv:2306.11632}\
  } (\bibinfo {year} {2023})}\BibitemShut {NoStop}%
\bibitem [{\citenamefont {Hashimoto}\ \emph {et~al.}(2023)\citenamefont
  {Hashimoto}, \citenamefont {Murata}, \citenamefont {Tanahashi},\ and\
  \citenamefont {Watanabe}}]{hashimoto2023krylov}%
  \BibitemOpen
  \bibfield  {author} {\bibinfo {author} {\bibfnamefont {K.}~\bibnamefont
  {Hashimoto}}, \bibinfo {author} {\bibfnamefont {K.}~\bibnamefont {Murata}},
  \bibinfo {author} {\bibfnamefont {N.}~\bibnamefont {Tanahashi}}, \ and\
  \bibinfo {author} {\bibfnamefont {R.}~\bibnamefont {Watanabe}},\ }\href@noop
  {} {\bibfield  {journal} {\bibinfo  {journal} {arXiv preprint
  arXiv:2305.16669}\ } (\bibinfo {year} {2023})}\BibitemShut {NoStop}%
\bibitem [{\citenamefont {Scialchi}\ \emph {et~al.}(2023)\citenamefont
  {Scialchi}, \citenamefont {Roncaglia},\ and\ \citenamefont
  {Wisniacki}}]{Diego}%
  \BibitemOpen
  \bibfield  {author} {\bibinfo {author} {\bibfnamefont {G.}~\bibnamefont
  {Scialchi}}, \bibinfo {author} {\bibfnamefont {A.}~\bibnamefont {Roncaglia}},
  \ and\ \bibinfo {author} {\bibfnamefont {D.~A.}\ \bibnamefont {Wisniacki}},\
  }\href@noop {} {\bibfield  {journal} {\bibinfo  {journal} {submitted}\ }
  (\bibinfo {year} {2023})}\BibitemShut {NoStop}%
\bibitem [{\citenamefont {García-Mata}\ \emph {et~al.}(2023)\citenamefont
  {García-Mata}, \citenamefont {Jalabert},\ and\ \citenamefont
  {Wisniacki}}]{García-Mata:2023}%
  \BibitemOpen
  \bibfield  {author} {\bibinfo {author} {\bibfnamefont {I.}~\bibnamefont
  {García-Mata}}, \bibinfo {author} {\bibfnamefont {R.~A.}\ \bibnamefont
  {Jalabert}}, \ and\ \bibinfo {author} {\bibfnamefont {D.~A.}\ \bibnamefont
  {Wisniacki}},\ }\href {\doibase 10.4249/scholarpedia.55237} {\bibfield
  {journal} {\bibinfo  {journal} {Scholarpedia}\ }\textbf {\bibinfo {volume}
  {18}},\ \bibinfo {pages} {55237} (\bibinfo {year} {2023})},\ \bibinfo {note}
  {revision \#199677}\BibitemShut {NoStop}%
\bibitem [{\citenamefont {Toscano}\ and\ \citenamefont
  {Wisniacki}(2006)}]{PhysRevE.74.056208}%
  \BibitemOpen
  \bibfield  {author} {\bibinfo {author} {\bibfnamefont {F.}~\bibnamefont
  {Toscano}}\ and\ \bibinfo {author} {\bibfnamefont {D.~A.}\ \bibnamefont
  {Wisniacki}},\ }\href {\doibase 10.1103/PhysRevE.74.056208} {\bibfield
  {journal} {\bibinfo  {journal} {Phys. Rev. E}\ }\textbf {\bibinfo {volume}
  {74}},\ \bibinfo {pages} {056208} (\bibinfo {year} {2006})}\BibitemShut
  {NoStop}%
\bibitem [{\citenamefont {Oganesyan}\ and\ \citenamefont
  {Huse}(2007)}]{oganesyan2007localization}%
  \BibitemOpen
  \bibfield  {author} {\bibinfo {author} {\bibfnamefont {V.}~\bibnamefont
  {Oganesyan}}\ and\ \bibinfo {author} {\bibfnamefont {D.~A.}\ \bibnamefont
  {Huse}},\ }\href@noop {} {\bibfield  {journal} {\bibinfo  {journal} {Phys.
  Rev. B}\ }\textbf {\bibinfo {volume} {75}},\ \bibinfo {pages} {155111}
  (\bibinfo {year} {2007})}\BibitemShut {NoStop}%
\bibitem [{\citenamefont {Kudo}\ and\ \citenamefont
  {Deguchi}(2018)}]{kudo2018finite}%
  \BibitemOpen
  \bibfield  {author} {\bibinfo {author} {\bibfnamefont {K.}~\bibnamefont
  {Kudo}}\ and\ \bibinfo {author} {\bibfnamefont {T.}~\bibnamefont {Deguchi}},\
  }\href@noop {} {\bibfield  {journal} {\bibinfo  {journal} {Phys. Rev. B}\
  }\textbf {\bibinfo {volume} {97}},\ \bibinfo {pages} {220201} (\bibinfo
  {year} {2018})}\BibitemShut {NoStop}%
\bibitem [{\citenamefont {Atas}\ \emph {et~al.}(2013)\citenamefont {Atas},
  \citenamefont {Bogomolny}, \citenamefont {Giraud},\ and\ \citenamefont
  {Roux}}]{atas2013distribution}%
  \BibitemOpen
  \bibfield  {author} {\bibinfo {author} {\bibfnamefont {Y.}~\bibnamefont
  {Atas}}, \bibinfo {author} {\bibfnamefont {E.}~\bibnamefont {Bogomolny}},
  \bibinfo {author} {\bibfnamefont {O.}~\bibnamefont {Giraud}}, \ and\ \bibinfo
  {author} {\bibfnamefont {G.}~\bibnamefont {Roux}},\ }\href@noop {} {\bibfield
   {journal} {\bibinfo  {journal} {Phys. Rev. Lett.}\ }\textbf {\bibinfo
  {volume} {110}},\ \bibinfo {pages} {084101} (\bibinfo {year}
  {2013})}\BibitemShut {NoStop}%
\bibitem [{\citenamefont {Bohigas}\ \emph {et~al.}(1984)\citenamefont
  {Bohigas}, \citenamefont {Giannoni},\ and\ \citenamefont
  {Schmit}}]{Bohigas1984}%
  \BibitemOpen
  \bibfield  {author} {\bibinfo {author} {\bibfnamefont {O.}~\bibnamefont
  {Bohigas}}, \bibinfo {author} {\bibfnamefont {M.~J.}\ \bibnamefont
  {Giannoni}}, \ and\ \bibinfo {author} {\bibfnamefont {C.}~\bibnamefont
  {Schmit}},\ }\href {\doibase 10.1103/physrevlett.52.1} {\bibfield  {journal}
  {\bibinfo  {journal} {Phys. Rev. Lett.}\ }\textbf {\bibinfo {volume} {52}},\
  \bibinfo {pages} {1} (\bibinfo {year} {1984})}\BibitemShut {NoStop}%
\bibitem [{\citenamefont {Chen}\ \emph
  {et~al.}(2020{\natexlab{b}})\citenamefont {Chen}, \citenamefont {Nurdin},\
  and\ \citenamefont {Yamamoto}}]{QRC}%
  \BibitemOpen
  \bibfield  {author} {\bibinfo {author} {\bibfnamefont {J.}~\bibnamefont
  {Chen}}, \bibinfo {author} {\bibfnamefont {H.~I.}\ \bibnamefont {Nurdin}}, \
  and\ \bibinfo {author} {\bibfnamefont {N.}~\bibnamefont {Yamamoto}},\
  }\href@noop {} {\bibfield  {journal} {\bibinfo  {journal} {Phys. Rev. Appl.}\
  }\textbf {\bibinfo {volume} {14}},\ \bibinfo {pages} {024065} (\bibinfo
  {year} {2020}{\natexlab{b}})}\BibitemShut {NoStop}%
\bibitem [{\citenamefont {Gottesman}(1998)}]{G2}%
  \BibitemOpen
  \bibfield  {author} {\bibinfo {author} {\bibfnamefont {D.}~\bibnamefont
  {Gottesman}},\ }\href@noop {} {\enquote {\bibinfo {title} {The {Heisenberg}
  representation of quantum computers},}\ } (\bibinfo {year} {1998}),\ \Eprint
  {http://arxiv.org/abs/quant-ph/9807006} {arXiv:quant-ph/9807006} \BibitemShut
  {NoStop}%
\bibitem [{\citenamefont {Jozsa}\ and\ \citenamefont {den Nest}(2014)}]{G1}%
  \BibitemOpen
  \bibfield  {author} {\bibinfo {author} {\bibfnamefont {R.}~\bibnamefont
  {Jozsa}}\ and\ \bibinfo {author} {\bibfnamefont {M.~V.}\ \bibnamefont {den
  Nest}},\ }\href@noop {} {\bibfield  {journal} {\bibinfo  {journal} {Quantum
  Inf. Comput.}\ }\textbf {\bibinfo {volume} {14}},\ \bibinfo {pages} {633}
  (\bibinfo {year} {2014})}\BibitemShut {NoStop}%
\bibitem [{\citenamefont {Jozsa}\ and\ \citenamefont
  {Miyake}(2008)}]{matchgates1}%
  \BibitemOpen
  \bibfield  {author} {\bibinfo {author} {\bibfnamefont {R.}~\bibnamefont
  {Jozsa}}\ and\ \bibinfo {author} {\bibfnamefont {A.}~\bibnamefont {Miyake}},\
  }\href@noop {} {\bibfield  {journal} {\bibinfo  {journal} {Proc. Math. Phys.
  Eng. Sci.}\ }\textbf {\bibinfo {volume} {464}},\ \bibinfo {pages} {3089}
  (\bibinfo {year} {2008})}\BibitemShut {NoStop}%
\bibitem [{\citenamefont {Brod}\ and\ \citenamefont
  {Childs}(2014)}]{matchgates2}%
  \BibitemOpen
  \bibfield  {author} {\bibinfo {author} {\bibfnamefont {D.~J.}\ \bibnamefont
  {Brod}}\ and\ \bibinfo {author} {\bibfnamefont {A.~M.}\ \bibnamefont
  {Childs}},\ }\href@noop {} {\bibfield  {journal} {\bibinfo  {journal}
  {Quantum Inf. Comput.}\ }\textbf {\bibinfo {volume} {14}},\ \bibinfo {pages}
  {901–916} (\bibinfo {year} {2014})}\BibitemShut {NoStop}%
\bibitem [{\citenamefont {Bremner}\ \emph {et~al.}(2011)\citenamefont
  {Bremner}, \citenamefont {Jozsa},\ and\ \citenamefont
  {Shepherd}}]{diagonals}%
  \BibitemOpen
  \bibfield  {author} {\bibinfo {author} {\bibfnamefont {M.~J.}\ \bibnamefont
  {Bremner}}, \bibinfo {author} {\bibfnamefont {R.}~\bibnamefont {Jozsa}}, \
  and\ \bibinfo {author} {\bibfnamefont {D.~J.}\ \bibnamefont {Shepherd}},\
  }\href@noop {} {\bibfield  {journal} {\bibinfo  {journal} {Proc. Math. Phys.
  Eng. Sci.}\ }\textbf {\bibinfo {volume} {467}},\ \bibinfo {pages} {459}
  (\bibinfo {year} {2011})}\BibitemShut {NoStop}%
\bibitem [{\citenamefont {Park}\ and\ \citenamefont
  {Light}(1986)}]{park1986unitary}%
  \BibitemOpen
  \bibfield  {author} {\bibinfo {author} {\bibfnamefont {T.~J.}\ \bibnamefont
  {Park}}\ and\ \bibinfo {author} {\bibfnamefont {J.}~\bibnamefont {Light}},\
  }\href {https://aip.scitation.org/doi/abs/10.1063/1.451548} {\bibfield
  {journal} {\bibinfo  {journal} {J. Chem. Phys.}\ }\textbf {\bibinfo {volume}
  {85}},\ \bibinfo {pages} {5870} (\bibinfo {year} {1986})}\BibitemShut
  {NoStop}%
\bibitem [{\citenamefont {Saad}(1992)}]{saad1992analysis}%
  \BibitemOpen
  \bibfield  {author} {\bibinfo {author} {\bibfnamefont {Y.}~\bibnamefont
  {Saad}},\ }\href {https://epubs.siam.org/doi/abs/10.1137/0729014?mobileUi=0}
  {\bibfield  {journal} {\bibinfo  {journal} {SIAM J. Numer. Anal.}\ }\textbf
  {\bibinfo {volume} {29}},\ \bibinfo {pages} {209} (\bibinfo {year}
  {1992})}\BibitemShut {NoStop}%
\bibitem [{\citenamefont {Stewart}\ and\ \citenamefont
  {Leyk}(1996)}]{stewart1996error}%
  \BibitemOpen
  \bibfield  {author} {\bibinfo {author} {\bibfnamefont {D.~E.}\ \bibnamefont
  {Stewart}}\ and\ \bibinfo {author} {\bibfnamefont {T.}~\bibnamefont {Leyk}},\
  }\href {https://epubs.siam.org/doi/10.1137/S0036142995280572} {\bibfield
  {journal} {\bibinfo  {journal} {J. Comput. Appl. Math.}\ }\textbf {\bibinfo
  {volume} {72}},\ \bibinfo {pages} {359} (\bibinfo {year} {1996})}\BibitemShut
  {NoStop}%
\bibitem [{\citenamefont {Hochbruck}\ and\ \citenamefont
  {Lubich}(1997{\natexlab{b}})}]{hochbruck1997krylov}%
  \BibitemOpen
  \bibfield  {author} {\bibinfo {author} {\bibfnamefont {M.}~\bibnamefont
  {Hochbruck}}\ and\ \bibinfo {author} {\bibfnamefont {C.}~\bibnamefont
  {Lubich}},\ }\href {https://doi.org/10.1137/S0036142995280572} {\bibfield
  {journal} {\bibinfo  {journal} {SIAM J. Numer. Anal.}\ }\textbf {\bibinfo
  {volume} {34}},\ \bibinfo {pages} {1911} (\bibinfo {year}
  {1997}{\natexlab{b}})}\BibitemShut {NoStop}%
\bibitem [{\citenamefont {Sidje}(1998)}]{expokit}%
  \BibitemOpen
  \bibfield  {author} {\bibinfo {author} {\bibfnamefont {R.~B.}\ \bibnamefont
  {Sidje}},\ }\href {\doibase 10.1145/285861.285868} {\bibfield  {journal}
  {\bibinfo  {journal} {ACM Trans. Math. Softw.}\ }\textbf {\bibinfo {volume}
  {24}},\ \bibinfo {pages} {130–156} (\bibinfo {year} {1998})}\BibitemShut
  {NoStop}%
\bibitem [{\citenamefont {Moler}\ and\ \citenamefont
  {Van~Loan}(2003)}]{moler2003nineteen}%
  \BibitemOpen
  \bibfield  {author} {\bibinfo {author} {\bibfnamefont {C.}~\bibnamefont
  {Moler}}\ and\ \bibinfo {author} {\bibfnamefont {C.}~\bibnamefont
  {Van~Loan}},\ }\href {https://doi.org/10.1137/S00361445024180} {\bibfield
  {journal} {\bibinfo  {journal} {SIAM Rev.}\ }\textbf {\bibinfo {volume}
  {45}},\ \bibinfo {pages} {3} (\bibinfo {year} {2003})}\BibitemShut {NoStop}%
\bibitem [{\citenamefont {Jawecki}\ \emph {et~al.}(2020)\citenamefont
  {Jawecki}, \citenamefont {Auzinger},\ and\ \citenamefont
  {Koch}}]{Jawecki:2020cc}%
  \BibitemOpen
  \bibfield  {author} {\bibinfo {author} {\bibfnamefont {T.}~\bibnamefont
  {Jawecki}}, \bibinfo {author} {\bibfnamefont {W.}~\bibnamefont {Auzinger}}, \
  and\ \bibinfo {author} {\bibfnamefont {O.}~\bibnamefont {Koch}},\ }\href
  {https://link.springer.com/10.1007/s10543-019-00771-6} {\bibfield  {journal}
  {\bibinfo  {journal} {BIT Numer. Math.}\ }\textbf {\bibinfo {volume} {60}},\
  \bibinfo {pages} {157} (\bibinfo {year} {2020})}\BibitemShut {NoStop}%
\end{thebibliography}%

\clearpage
\newpage
\begin{widetext}
\section*{Supplemental Material for \\ \textit{Quantum reservoir complexity by Krylov evolution approach}}
\subsection{Families of Quantum reservoirs}
\label{sect:QR}

The seven families of quantum reservoirs considered in this work are the following: 

\begin{enumerate}
    \item \textbf{G1:} The quantum circuit is constructed from the generator G1 = \{CNOT, H, X\}, where CNOT is the controlled-NOT gate, H stands for Hadamard, and X is the NOT gate. The set G1 generates a subgroup of Clifford \cite{G1} group, and thus is non-universal and classically simulatable.
    \item \textbf{G2:} The quantum circuit is constructed from the generator G2 = \{CNOT, H, S\}, with S being the $\pi/4$ phase gate. The circuits constructed from G2 generate the whole Clifford group \cite{G2}, so they are non-universal and classically simulatable, but more complex than G1 circuits.
    \item \textbf{G3:} The quantum circuit is constructed from the generator 
    G3=\{CNOT,H,T\}, where $T$ is the $\pi/8$ phase gate. G3 is universal and thus 
    approximates the full unitary group $U(N)$ to arbitrary precision.
    \item \textbf{Matchgates (MG):} Two-qubit gates formed by 2 one-qubit gates, 
    $A$ and $B$, with the same determinant. $A$ acts on the subspace spanned by $\ket{00}$ and $\ket{11}$, while $B$ acts on the subspace  spanned by $\ket{01}$ and $\ket{10}$. $A$ and $B$ are randomly sampled from the unitary group U(2):
    \begin{equation}
        G(A,B) = 
        \begin{pmatrix} 
            a_1 & 0 & 0 & a_2 \\
            0 & b_1 & b_2 & 0 \\
            0 & b_3 & b_4 & 0 \\
            a_3 & 0 & 0 & a_4
        \end{pmatrix}, \quad |A| = |B|.
    \end{equation}
    Matchgates circuits are also universal (except when acting only on nearest neighboring lines) \cite{matchgates1, matchgates2}. 
    \item \textbf{Diagonal-gate circuits ($\mathbf{D_2}$, $\mathbf{D_3}$, $\mathbf{D_n}$):} The last families of gates are diagonal in the computational basis. The diagonal gates are separated into 3 families: $D_2$, $D_3$ and $D_n$. Here, $D_2$ gates are applied to pairs of qubits, $D_3$ gates are applied to 3 qubits, and $D_n$ gates are applied to all qubits.
    \begin{equation}
        D_k(\phi_1, \cdots, \phi_{2^k}) = 
        \begin{pmatrix} 
            e^{i\phi_1} & 0 & \cdots & 0  \\
            0 & e^{i\phi_2} & \cdots & 0 \\
            \vdots & \vdots  & \ddots & \vdots \\
             0 & 0 & \cdots & e^{i\phi_{2^k}}
        \end{pmatrix}, 
    \end{equation}
    for $k \in \{2,3,n\}$, and with $\phi_i$ chosen uniformly from $[0, 2\pi) \ \forall i$. The gates are applied on all combinations of $k$ (out of $n$) qubits, the ordering being random. At the beginning and at the end of the circuit (after the initialization of the state), Hadamard gates are applied to all qubits. As diagonal gates commute, they can be applied simultaneously. Diagonal circuits cannot perform universal computation but they are not always classically simulatable \cite{diagonals}. As opposed to the other families of circuits, which can be of arbitrary depth, the diagonal $D_2$, $D_3$ and $D_n$ families contain a fixed number of gates, being those $\binom{n}{2}$, $\binom{n}{3}$ and 1 gates, respectively. 
\end{enumerate}

\subsection{Krylov approach with reduced Lanczos coefficients}
\label{sect:reduced_Lanczos}

\begin{figure}
    \centering
    \includegraphics[width=0.5\columnwidth]{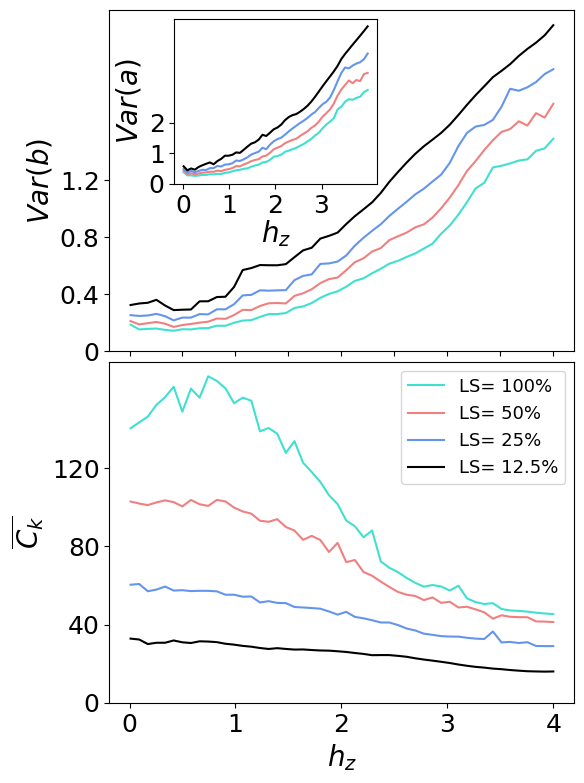}
    \caption{(top) Variance of the Lanczos coefficients $a$ and $b$ for the Ising model with $n = 10$ spins, 
    as a function of $h_z$  
    (bottom) Mean K-complexity ($\overline{C_k}$) as a function of $h_z$. 
    The size of the Lanczos sequence (LS) is reduced to 100\%,  50\%, 25\% and 12.5\%. }
    \label{fig:short-H}
\end{figure}

In this section, we study the performance of the Krylov approach when considering only a simplified 
subset of coefficients from the complete Lanczos sequence. 
Rather than calculating the Krylov complexity and statistics using all coefficients $a$ and $b$, 
we perform the calculations here using only a fraction of these coefficients while disregarding 
the tail end of the sequence. 
The percentage of the sequence of the Lanczos coefficients used for the calculations will be called 
Lanczos Sequence (LS) from now on.

Figure~\ref{fig:short-H} (top panel) illustrates the variance of $a$ and $b$, 
while Fig.~\ref{fig:short-H} (bottom panel) shows the mean complexity $\overline{C_k}$ 
for the Ising model with $n=10$ spins. 
The colored lines in the graphs represent the obtained statistics using all Lanczos coefficients 
(green) and also when using only a fraction of them, with LS = 50\% (red), 25\% (blue), 
and 12.5\% (black). 
As can be seen in the bottom panel, the transition of the $K-$complexity is accurately reproduced when using all Lanczos coefficients and also fairly well reproduced when only 50\% of the coefficients are included. 
On the contrary, the $K-$complexity phase transition is completely lost when using only a small 
fraction of coefficients, for example, 12.5\%. 
On the other hand (top panel), the evolution of the variance of the Lanczos coefficients 
is well reproduced even when using very small fractions of the Lanczos sequence, 
as it consistently exhibits a growing behavior with respect to $h_z$. 
More interestingly, the variance obtained from smaller fractions of coefficients actually shows 
higher variances than those obtained from the entire set of coefficients. 
These findings indicate that when studying the dynamics of Lanczos coefficients, one can focus solely 
on the first few values and still achieve accurate results, leading to a significant reduction in the computational complexity of the method. 
As for the $K-$complexity, its evolution is reasonably replicated when using smaller sets of coefficients, 
but its trend becomes completely lost when considering only the initial few values.

\begin{figure}[t!]
    \centering
    \includegraphics[width=0.5\columnwidth]{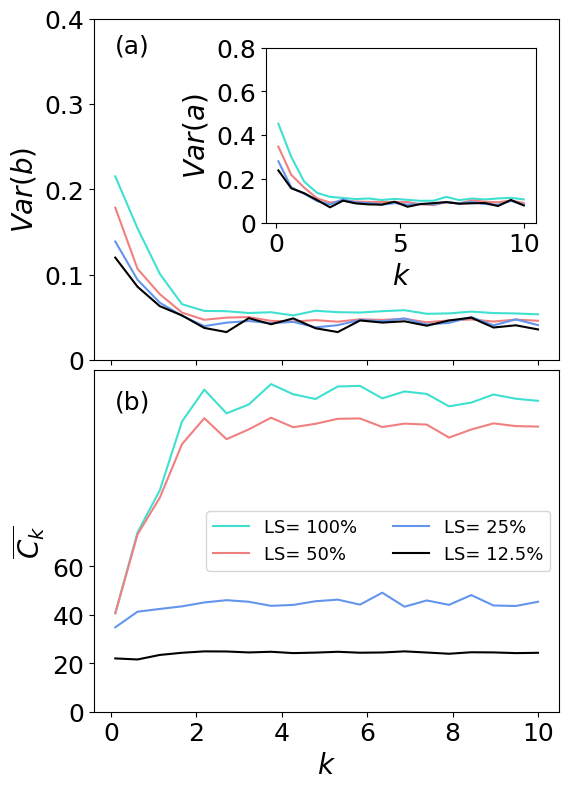}
    \caption{Same as Fig. \ref{fig:short-H} for the quantum standard map with $N=400$. }
    \label{fig:short-U}
\end{figure}

Finally, the same trends are observed when studying the quantum standard map for $N=400$, 
as depicted in Fig.~\ref{fig:short-U}.
The behavior of the variance of the Lanczos coefficients is well-reproduced even for smaller sets 
of coefficients, with as little as 12.5\% of them being sufficient. 
Moreover, the $K-$complexity is also well-reproduced when using only 50\% of the coefficients, 
but its dynamics are lost when using 25\% and 12.5\% of the coefficients. 
These results highlight the computational efficiency of the Krylov approach, 
as significantly reducing the size of the Lanczos coefficients still allows for an accurate
reproduction of the Krylov statistics and complexity. 
Consequently, employing only the first coefficients of the Lanczos sequence can substantially reduce 
the computational time and resources required to apply this method to large-scale systems.
\end{widetext}

\end{document}